\documentclass[%
reprint,
amsmath,amssymb,
pra,
tightenlines
]{revtex4-1}

\usepackage{graphicx}
\usepackage{dcolumn}
\usepackage{bm}
\usepackage{color}
\usepackage{graphicx}
\usepackage{float}
\usepackage{amsmath}
\graphicspath{ {images/} }

\newcommand{\rvec}{\mathbf{r}} 
\newcommand{\Hmat}{\mathbf{H}} 
\newcommand{\Lmat}{\mathbf{L}} 
\newcommand{\Nmat}{\mathbf{N}} 
\newcommand{\Vmat}{\mathbf{V}} 
\newcommand{\Umat}{\mathbf{U}} 
\newcommand{\rev}[1]{{\color{black} #1}}

\begin{document}

\title{Exponential Integrators in Time-Dependent Density
Functional Calculations}


\author{Daniel Kidd}
\author{Cody Covington}%
\author{K\'alm\'an Varga}
 \email{kalman.varga@vanderbilt.edu}
\affiliation{%
 Department of Physics and Astronomy, Vanderbilt University, Nashville, Tennessee 37235, USA
}%

\date{\today}

\begin{abstract}
The integrating factor and exponential time differencing methods are implemented and tested for
solving the time-dependent Kohn--Sham equations. Popular
time propagation methods used in physics, as well as other robust numerical
approaches, are compared to these exponential integrator methods in order to judge the relative
merit of the computational schemes. We determine an improvement in accuracy of multiple orders of
magnitude when describing dynamics driven primarily by a nonlinear potential. For cases of dynamics 
driven by a time-dependent external potential, the accuracy of the exponential integrator methods
are less enhanced but still match or outperform the best of the
conventional methods tested.  
\end{abstract}

\maketitle

\section{\label{sec:intro}Introduction}

Time-dependent density functional theory (TDDFT)
\cite{runge_prl_52_997_1984,ullrich_bjp_44_154_2014}
has become widely used in the simulation of molecules, two-dimensional materials,
and bulk materials. Applications of TDDFT include the calculation of optical
properties \cite{tsolakidis_prb_66_235416_2002,Tsolakidis2005,Johnson2009,tsolakidis_prb_66_235416_2002,Aikens2008,
doi:10.1063/1.3671952}, charge transfer dynamics
\cite{doi:10.1021/acs.jctc.5b00895,PhysRevA.83.042501}, excitations
\cite{BENDE2015103,meng_jcp_129_054110_2008},
field emission
\cite{driscoll_n_22_285702_2011,driscoll_prb_83_233405_2011,driscoll_jap_110_024304_2011,driscoll_prb_80_245431_2009},
ultrafast strong field processes \cite{chu_pra_64_063404_2001,bubin_pra_86_043407_2012,bubin_apl_98_154101_2011,
bubin_jap_110_064905_2011,bubin_prb_85_205441_2012,PhysRevA.91.023422,PhysRevA.92.053413,
haruyama_pra_85_062511_2012,heslar_pra_87_052513_2013,isla_jpcc_111_17765_2007,livshits_jpca_110_8443_2006},
and ion collisions \cite{krasheninnikov_prl_99_016104_2007,bubin_prb_85_235435_2012,hatcher_prl_100_103201_2008}.
Efficient computer codes for TDDFT calculations have been developed by
several groups \cite{Andrade2012,meng_jcp_129_054110_2008,Noda2014145,DRAEGER2017205,C7CP00995J}.

Many of these calculations require long time stability and accuracy,
but as we have shown in a recent paper \cite{PhysRevE.94.023314},
initial stability does not guarantee that the numerical solution does
not gradually deteriorate from the correct one. Part of the problem 
is due to the handling of the nonlinear, density-dependent part of the 
potential. In quantum mechanical calculations, this 
nonlinear part is time propagated using the time evolution operator 
together with the rest of the Hamiltonian. The only difference is that an extra
self-consistency step is added to each time step
\cite{meng_jcp_129_054110_2008,PhysRevB.59.2579} in order to ensure that the
density and the Hamiltonian are instantaneously self-consistent.

Various mathematical approaches have been developed for the solution of nonlinear
differential equations. In order to solve the initial value problem
\begin{equation}
{d y\over d  t}=f(y,t)\ \ \ \ \ \ y(t=0)=y_0,
\end{equation}
one can separate the linear, $Ly$, and nonlinear, $N(y,t)$, terms as
\begin{equation}
{d y\over d  t}=Ly+N(y,t).
\label{main}
\end{equation}
The way in which the linear and nonlinear terms each govern the solution
depends on the type of the operators and can be very different. The best
approach for solving the problem is to develop separate approximations
that are best suited to the linear and nonlinear part individually.

Three robust numerical methods
have been developed and tested for this purpose, the
implicit-explicit (IMEX) \cite{imex}, the integrating factor (IF) \cite{COX2002430}, and the
exponential time differencing (ETD) methods \cite{COX2002430}.
The latter two methods are collectively  known as exponential integrators. 
The IMEX multistep method solves the stiff linear part of the equation with an
implicit scheme and the nonlinear part with an explicit scheme. The implicit
approaches are more stable but computationally more demanding; the
explicit method is only conditionally stable. The IF
method introduces a new variable by factoring out the stiff part of the
equation, and only the nonlinear part has to be solved by time stepping.
In the ETD method, the
exact integration of the linear part is followed by an approximate
integration of the nonlinear part. These approaches have been tested
for dissipative and dispersive partial diffential equations; examples
include the Allen--Cahn, Burgers, Cahn--Hilliard, Kuramoto--Sivashinsky, Navier--Stokes, and 
Swift--Hohenberg equations
\cite{Boyd99chebyshevand,Fornberg99afast,COX2002430,doi:10.1137/S1064827502410633,doi:10.1137/S1064827597321532}. 
The ETD approach seems to be the most accurate  in test
calculations \cite{COX2002430,doi:10.1137/S1064827502410633}. 

In this paper we will implement and test the exponential integrator approaches in TDDFT calculations. 
There are three important distinctions between the differential equations solved in TDDFT
and the first-order nonlinear differential equations considered  in
the mathematical literature: (1) the coupled nature of the TDDFT equations, (2) the
time-dependent external potential, and (3) the Hartree potential. The density
calculated from the orbitals couples the TDDFT equations through the
nonlinear potential. The external potential is a 
time-dependent linear part of the differential
equation; no such term has been used in the
nonlinear ODE studies. The Hartree potential makes Eq. \eqref{main} an
integro-differential equation. In order to test the ETD and IF approaches, we will
compare them to the IMEX and
conventional time evolution schemes popular in physics. We will use
one-dimensional models which will allow for large spatial simulation boxes, avoiding
artificial reflection, and long time propagation for clean, stringent
comparison. At the same time, we expect that these test systems
present the same possible problems (nonlinearity, coupling) as one must face
in larger systems.

While the application of time evolution operators and split operator
representations date back to the 1950s, ETD
\cite{taf05,BEYLKIN1998362,COX2002430,doi:10.1137/S1064827502410633} and IF
\cite{Krogstad:2005:GIF:1063068.1063072,doi:10.1137/S1064827502410633} methods
are relatively new. That may explain why they are not used
in time-dependent quantum mechanical calculations. We hope that the
examples presented in this paper will pave the way for the application
of exponential integrator approaches in TDDFT calculations. 

\section{Formalism}
Consider the time-dependent Kohn--Sham (TDKS) equations
\begin{align}\label{eq:tddft}
i{\partial \over \partial t} \Psi({\mathbf{r},t})=&\Hmat({\mathbf{r},t})\Psi({\mathbf{r},t})\\ \notag =& 
\Lmat({\mathbf{r},t})\Psi({\mathbf{r},t})+\Nmat({\Psi,t}),
\end{align}
where  $\left\{\Psi=\Psi^1,\Psi^2,{\ldots}\right\} $ is the set of Kohn--Sham orbitals, 
\begin{equation}
\Lmat=\mathbf{T}+\Vmat
\end{equation}
is the linear part, comprised of kinetic matrix, $\mathbf{T}$, and linear, time-independent potential, 
$\Vmat$, and
\begin{equation}
\Nmat(\Psi,t)=\Vmat^\text{N}(\Psi,t) \Psi({\mathbf{r},t})
\end{equation}
is the nonlinear part of the Hamiltonian. The nonlinear part depends
on all orbitals, coupling the differential equations. The nonlinear potential, $\Vmat^\text{N}$, is the sum of the
Hartree and exchange-correlation contributions, plus the time-dependent
potential. The latter is a linear term, but in the rest of the
formalism it is more convenient to absorb it into $\Vmat^\text{N}$, keeping $\Lmat$
time-independent.

\subsection{Time evolution operator approach}
The formal solution of Eq. \eqref{eq:tddft} can be obtained by using the
time evolution operator
\begin{equation}
\Psi({\mathbf{r},t})=\Umat(t,0)\Psi({\mathbf{r},0}).
\end{equation}
The time evolution operator is defined as
\begin{equation}
\label{exact_time_ev_operator}
\Umat(t,0)=\mathcal{T} \exp \left[-\frac{i}{\hbar} \int \Hmat(\mathbf{r},t') d
t' \right], 
\end{equation}
where $\mathcal{T}$ denotes time-ordering. Two important properties of
the time evolution operator are the following: (1) it is unitary for Hermitian
Hamiltonians and (2) it has time reversal symmetry. 

In practice, the above
expression for $\Umat(t,0)$
is split into a product of multiple time evolution operators, each
corresponding to a short
time step $\Delta t$, 
\begin{equation}
\Umat(0,t)=\prod_{n} \Umat(t_n, t_{n+1}), \quad t_n= n \Delta t,
\end{equation}
\begin{equation} 
\Umat(t_n, t_{n+1}) =\exp \left[-\frac{i}{\hbar} \Hmat(t_n) \Delta
t \right],
\label{prop_dt}
\end{equation}
so that the Hamiltonian at 
time $t_n$ remains nearly commutative with the Hamiltonian at time
$t_{n+1}$. 
Various schemes have been developed for the construction of the time propagator
\cite{PhysRevE.73.036708,1.4921465,1.3549570,PhysRevE.70.056703,1.3126363,1.475495,
Kosloff198335,FEIT1982412,1.480347}
including polynomial propagators \cite{timeprop,tal-ezer_jcp_81_3967_1984},
exponential propagators \cite{S0219633613400014,Bandrauk2006346},
subspace propagation \cite{ParkLight1986,Manthe1991,leforestier_jcp_94_59_1991,Manthe1998442,
Smyth19981,Farnum2004,isla_jpcc_111_17765_2007,PhysRevLett.108.163001,PhysRevA.89.023418,
Lehner2015,PhysRevB.82.205410,Wang2015,timeprop},
and split-operator techniques \cite{timeprop}. 
The overwhelming majority of TDDFT and other time-dependent
Schr\"odinger equation based calculations use the time evolution
approach and only differ in the representation of the exponential
operator. In these approaches, the linear and nonlinear parts are not
separated, and the separation of slowly and rapidly varying parts is not
exploited. 

\subsection{Implicit-explicit schemes}
Various IMEX schemes \cite{imex} have been developed to solve nonlinear
differential equations. In these approaches the linear part is
advanced with an implicit scheme---for example, with the Adams--Moulton
method (see Eq. \eqref{am2})---while the nonlinear part is handled with an explicit multistep
method---for example, the Adams--Bashforth formula (see Eq. \eqref{ab2}). The implicit scheme
is stable, so larger time steps may be used; the explicit scheme is only
conditionally stable and requires smaller time steps. Appendix \ref{app:AM} gives a
brief summary of these approaches. 

\subsection{Integrating Factor method}
The idea of the integrating factor approach is to multiply the 
differential equation by some integrating factor, thereby introducing new
variables. Ideally, one changes the variables to solve the linear part
exactly and uses some technique to solve the remaining nonlinear part. 
In the context of the TDKS equations, one may define
\begin{equation}
\Phi({\mathbf{r},t})={\rm e}^{-i\Lmat t}\Psi({\mathbf{r},t}),
\end{equation}
where the integrating factor is defined as ${\rm e}^{i\Lmat t}$ \cite{Boyd99chebyshevand,Fornberg99afast,COX2002430,
doi:10.1137/S1064827502410633,doi:10.1137/S1064827597321532,Krogstad:2005:GIF:1063068.1063072}.
By multiplying Eq. \eqref{eq:tddft} by the
integrating factor, one has
\begin{equation}
i{\partial \over \partial t} \Phi({\mathbf{r},t})=
{\rm e}^{-i\Lmat t}\Nmat({{\rm e}^{i\Lmat t}\Phi,t}).
\label{if}
\end{equation}
This approach is closely related to the interaction picture in quantum
mechanics. The aim of the transformation is to ameliorate the stiff
linear part of the differential equation. One can then use a time
stepping method, for example Runge--Kutta  or Adams--Bashforth formulae, 
to advance the equation in time.  The disadvantage of the method is
that it changes the fixed points of the original differential equation, and the
local truncation error is larger than in other methods such as ETD
\cite{COX2002430}.

\subsection{Exponential time differencing}
The exponential time differencing method
has been used as early as in the 1960s for ordinary differential
equations \cite{Pope,10.2307/2949405}. It surfaced in
computational electrodynamics in the 1990s \cite{taf05} and was
rediscovered for numerical solutions of nonlinear differential equations later
\cite{BEYLKIN1998362,COX2002430,hochbruck_ostermann_2010,LUAN2014168,TOKMAN20118762}.

Using the identity
\begin{equation}\label{eq:etd_iden}
i{\partial \over \partial t}
\left[{\rm e}^{i\Lmat t}\Psi({\mathbf{r},t})\right]
=
{\rm e}^{i\Lmat t}\left[ -\Lmat\Psi({\mathbf{r},t})
+
i{\partial \over \partial t} \Psi({\mathbf{r},t})
\right],
\end{equation}
we can rewrite Eq. \eqref{eq:tddft} as
\begin{equation}
i{\partial \over \partial t}
\left[{\rm e}^{i\Lmat t}\Psi({\mathbf{r},t})\right]=
{\rm e}^{i\Lmat t}\Nmat({\Psi,t}).
\end{equation}
By integrating this equation from $t_n$ to $t_{n+1}$ and rearranging the
terms, one arrives at
\begin{multline}
\Psi(\mathbf{r},t_{n+1})={\rm e}^{-i\Lmat\Delta t}\Psi(\mathbf{r},t_n)\\
-i{\rm e}^{i\Lmat (t_{n+1})}\int_{t_n}^{t_{n+1}} 
{\rm e}^{i\Lmat\tau} \Nmat(\Psi,\tau) d \tau.
\label{etd}
\end{multline}
This equation is exact.  The difference between the ETD and IF
methods is that for ETD, the variable change is not complete---one
keeps $\Psi$ as the variable.

In the above derivation, it has been assumed that the linear term, $\Lmat$,
is time-independent and that all time-dependent terms, including those 
that are linear, have been incorporated within $\Nmat({\Psi,t})$. However one
may, instead, include linear time-dependent terms, $\Vmat^L(t)$, within $\Lmat$ and arrive
at the same conclusion as eq. \eqref{etd} if it can be assumed that
$\Vmat^L(t_n)\approx\Vmat^L(t_{n+1})$. This is shown in appendix \ref{app:lin_time}.

In practice, one solves the integral appearing in eq. \eqref{etd} by using some
approximation. In the evaluation of the integral, matrix-valued 
functions arise (e.g. $f(\Lmat)=\Lmat^{-1}$), as will be shown below. These
functions, together with matrix exponentials, must be
evaluated efficiently for applications \cite{LU2003203,doi:10.1137/S0036142995280572}. 
One can calculate the
exponential by Taylor expansion and obtain other needed functions 
by recurrence relations \cite{BEYLKIN1998362}. 
Krylov subspace methods seem to be optimal for large
matrices \cite{Tokman:2006:EIL:1145976.1145991}. In this case, the 
matrix functions are efficiently evaluated in a Krylov subspace, a process similar to
Lanczos time-propagation \cite{jcp/144/20/10.1063/1.4952646}.

The Cauchy formula
\begin{equation}
f(\Lmat)={1\over 2\pi i}\int_\Gamma f(t)(t\mathbf{I}-\Lmat)^{-1} dt, 
\end{equation}
where $\mathbf{I}$ is the identity matrix, as suggested in 
Ref. \cite{doi:10.1137/S1064827502410633}, can also be used to
evaluate these functions. The advantage of this approach is that by using
the trapezoidal rule on a complex contour, numerical instabilities
arising from possible small eigenvalues may be avoided. This method
can also be implemented within the Krylov subspace approach by
defining $\Lmat$ on the Krylov vectors \cite{Niesen:2012:A9K:2168773.2168781}. 
Accurate rational approximations \cite{Schmelzer:2006:EMF} and polynomial representations 
\cite{suhov} have also been developed.
Note that if $\Lmat$ is time-independent, the calculation of the exponential and
other functions of $\Lmat$ must only be done once.

\section{Solution of the TDDFT equation}
In this section we describe prototypical approaches for solving the TDKS
equations. The first three approaches---Taylor, split operator, and
Crank--Nicolson---are time evolution operator approaches and are widely 
used in time-dependent
quantum mechanical calculations. The IMEX, IF, and ETD methods 
have not been tested for TDDFT. In these
cases, we use the time propagating schemes developed in Refs. 
\cite{doi:10.1137/S1064827502410633,COX2002430}. These approaches are
simple to derive using  popular integration schemes (see Appendix \ref{app:methods}). 
A summary of the operation count, number of times the Hartree 
potential must be calculated per time step, and accuracy order with respect to 
time step size may be found in Table \ref{tab:method_summary}.

\begin{table}
\begin{tabular}{|c|c|c|c|}\hline
Method & Operations & Hartree & Accuracy \\\hline\hline
Taylor   & 4 & 1 &${\cal O}(\Delta t^4)$ \\\hline
SPO      & 1 & 1 &${\cal O}(\Delta t^3)$ \\\hline
CN       & 2 & 1 &${\cal O}(\Delta t^2)$ \\\hline
RK2      & 2 & 2 &${\cal O}(\Delta t^2)$ \\\hline
RK4      & 4 & 4 &${\cal O}(\Delta t^4)$ \\\hline\hline
AB2AM2   & 2 & 1 &${\cal O}(\Delta t^3)$ \\\hline\hline
IFAB2    & 3 & 1 &${\cal O}(\Delta t^3)$ \\\hline
IFRK2    & 3 & 2 &${\cal O}(\Delta t^3)$ \\\hline
IFRK4    & 3 & 4 &${\cal O}(\Delta t^4)$ \\\hline
ETD1     & 2 & 1 &${\cal O}(\Delta t^2)$ \\\hline
ETD2     & 3 & 1 &${\cal O}(\Delta t^3)$ \\\hline
ETDCN    & 3 & 1 &${\cal O}(\Delta t^3)$ \\\hline
ETDRK2   & 3 & 2 &${\cal O}(\Delta t^3)$ \\\hline
ETDRK4   & 8 & 4 &${\cal O}(\Delta t^4)$ \\\hline
Krogstad & 9 & 4 &${\cal O}(\Delta t^4)$ \\\hline
\end{tabular}
\caption{The main computational effort per time step 
is matrix vector multiplication (Operations) and solution of the Poisson equation (Hartree). 
All matrices dependent upon $\Lmat$ are considered to be constant in 
time so that they must be calculated only once. The table is separated into three sections: 
time evolution methods (top), in which the complete Hamiltonian is used to propagate the 
wave function, IMEX methods (middle), and exponential integrator methods (bottom), both of which split the 
Hamiltonian in into linear and nonlinear parts.}
\label{tab:method_summary}
\end{table}

\subsection{Taylor time propagation (Taylor)}
One propagation scheme of particular note is the fourth-order Taylor
propagation \cite{yabana_prb_54_4484_1996,bertsch_prb_62_7998_2000}.
In this scheme the exponential of the Hamiltonian, see
Eq.\,\eqref{prop_dt}, is approximated using a fourth-order Taylor expansion such that
\begin{equation}
\Psi(\mathbf{r},t_{n+1}) \approx
  \sum\limits_{k=1}^{4}\frac{1}{k!}\left(- \frac{i \Delta t}{\hbar}
  \Hmat(\mathbf{r},t_n) \right)^k \Psi(\mathbf{r},t_n).
  \label{taylor_propagator}
\end{equation}
Taylor propagation of the TDKS equations has proven highly successful
in many applications \cite{driscoll_n_22_285702_2011,driscoll_prb_83_233405_2011,
driscoll_jap_110_024304_2011,driscoll_prb_80_245431_2009,bubin_pra_86_043407_2012,
bubin_apl_98_154101_2011,bubin_jap_110_064905_2011,bubin_prb_85_205441_2012,
bubin_prb_85_235435_2012,PhysRevA.91.023422,PhysRevA.92.053413},
and its popularity is due to the simplicity of its implementation: only
matrix-vector multiplication is needed, while inversion and
diagonalization is avoided.  One drawback of this approach is that the
Taylor propagator is only conditionally stable.  
One also notes that the Taylor approximation breaks the unitarity of
the propagator.

\subsection{Split operator time propagation (SPO)} \label{sec:spo}
The split operator approach has a long history, first appearing in
Ref. \cite{split1} and independently as ``Sprang splitting'' in Ref. 
\cite{doi:10.1137/0705041}. The idea is to split the Hamiltonian into
kinetic and potential energy parts and approximate the propagator with 
a product of the exponentials of these operators.
This approach was first used in physics in Ref. \cite{FEIT1982412}, and
it was later developed for TDDFT using higher order decompositions 
in \cite{PhysRevB.59.2579}.

This approach is a derivative of the time evolution operator approach,
in which the discrete time step propagator, Eq. \eqref{prop_dt}, is 
approximated as
\begin{equation}
{\rm e}^{-i\Hmat_n\Delta t/\hbar}  \approx {\rm e}^{-i\mathbf{T}\Delta t/2\hbar}{\rm e}^{-i\Vmat_n\Delta t/\hbar}{\rm e}^{-i\mathbf{T}\Delta t/2\hbar}
\end{equation}
or, similarly,
\begin{equation}
{\rm e}^{-i\Hmat_n\Delta t/\hbar}\approx {\rm e}^{-i\Vmat_n\Delta t/2\hbar}{\rm e}^{-i\mathbf{T}\Delta t/\hbar}{\rm e}^{-i\Vmat_n\Delta t/2\hbar}.
\end{equation}
The above expressions are accurate to order ${\cal O}(\Delta t^3)$ \cite{Jahnke2000}.
Such splitting is chosen so that each matrix exponential is diagonal in either real space
or reciprocal space, facilitated by fast Fourier transforms. 
We note that this approach may also be applied to approximating the matrix 
exponential, $e^{-i\Lmat t}$, appearing in the IF method.

\subsection{Crank--Nicolson time propagation (CN)}
By adding the forward and backward Euler approaches (see Appendix \ref{app:Euler}), 
one gets the unconditionally stable Crank--Nicolson propagation scheme,
\begin{equation}\label{eq:CN}
y_{n+1}=y_n+{\Delta t\over 2} \left[
f(y_{n},t_{n})+f(y_{n+1},t_{n+1})\right],
\end{equation}
which is ${\cal O}(\Delta t^2)$ accurate in time.
Here,
\begin{multline} 
\left[\mathbf{I}+\frac{i\Delta t}{2}\left(\Lmat+\Vmat^\text{N}(\Psi_{n+1},t_{n+1})\right)\right] \Psi_{n+1} = \\ \left[\mathbf{I}-\frac{i\Delta t}{2}\left(\Lmat+\Vmat^\text{N}(\Psi_{n},t_{n})\right)\right] \Psi_n
\end{multline}
or, in the limit of small $\Delta t$ in which $\Vmat^\text{N}(\Psi_{n+1},t_{n+1})\approx \Vmat^\text{N}(\Psi_{n},t_{n})$,
\begin{multline} 
\Psi_{n+1} = \left[\mathbf{I}+\frac{i\Delta t}{2}\left(\Lmat+\Vmat^\text{N}(\Psi_{n},t_{n})\right)\right]^{-1} \\ 
\times \left[\mathbf{I}-\frac{i\Delta t}{2}\left(\Lmat+\Vmat^\text{N}(\Psi_{n},t_{n})\right)\right] \Psi_n.
\end{multline}
This method has the advantage of preserving the unitarity of the
time propagator. A disadvantage of the CN approach is the need
for the calculation of matrix inverses. While iterative calculations of
matrix inverses is possible for large sparse matrices, the application
of the CN method is not viable in grid based TDDFT calculations for
large systems.

\subsection{Second-order implicit-explicit scheme (AB2AM2)}
As an example to test an IMEX method, we use second-order
integration (see Appendix \ref{app:AM}). The linear terms will be handled using
the second-order Adams--Moulton method (trapezium
rule) and the nonlinear terms with the second-order 
Adams--Bashforth method. Adding Eqs. \eqref{am2} and \eqref{ab2} results in
\begin{eqnarray}
\Psi_{n+1}&=&\Psi_n-{i \Delta t\over 2} 
\left(\Lmat\Psi_n+\Lmat \Psi_{n+1}\right)\\
&-&
{3i\Delta t\over2} \Nmat(\Psi_n,t_n)
+{i\Delta t\over2} \Nmat(\Psi_{n-1},t_{n-1}).\nonumber
\end{eqnarray}
This can be solved for $\Psi_{n+1}$ as
\begin{eqnarray}
\Psi_{n+1}&=&
\left(\mathbf{I}+{i\Delta t\over 2} \Lmat\right)^{-1} \left[\left(\mathbf{I}-{i\Delta t\over
2} \Lmat\right) \Psi_n \right.
\\
&-&\left.{i\Delta t\over 2}
\left[3
\Nmat(\Psi_n,t_n)-\Nmat(\Psi_{n-1},t_{n-1})\right]\right]. \nonumber
\end{eqnarray}
This approach is ${\cal O}(\Delta t^3)$ accurate in time.

\subsection{Integrating factor method with explicit multistep (IFAB2)}
The integrating factor equation, Eq. \eqref{if}, can be solved using
the most popular integration schemes (see Appendix \ref{app:methods}). Using the
Adam-Bashforth method, as described in Eq. \eqref{ab2}, one has
\begin{multline}
\Psi_{n+1}={\rm e}^{i\Lmat \Delta t} \Psi_{n}+
{3 i \Delta t\over 2} {\rm e}^{i\Lmat \Delta t} \Nmat(\Psi_n,t_n) \\
-{i \Delta t\over 2} {\rm e}^{-2i\Lmat\Delta t} \Nmat(\Psi_{n-1},t_{n-1}).
\end{multline}
This approach is also ${\cal O}(\Delta t^3)$ accurate in time.

\subsection{Integrating factor method with second-order Runge--Kutta (IFRK2)}
By applying the second-order Runge--Kutta (RK2) method to Eq. \eqref{if} and
transforming the new variable, $\Phi$, back to $\Psi$,
one achieves a new time propagation update scheme of accuracy
${\cal O}(\Delta t^3)$,
\begin{multline}
\Psi_{n+1}={\rm e}^{i\Lmat \Delta t} \Psi_n \\ -{i\over\Delta t} \left(\Nmat(\Phi_a,t_n) +
{\rm e}^{i\Lmat \Delta t} \Nmat(\Psi_n,t_n)\right),
\end{multline}
where
\begin{equation}
\Phi_a= {\rm e}^{i\Lmat \Delta t} \Psi_{n}+\mathbf{M}_1 \Nmat(\Psi_{n},t_{n}).
\end{equation}

\subsection{Integrating factor method with fourth-order Runge--Kutta (IFRK4)}
The above approach can be extended for fourth-order Runge--Kutta (RK4) by preparing four
vectors 
\begin{eqnarray}
\Phi_1&=& \Nmat(\Psi_n,t_n),\\
\Phi_2&=& \Nmat\left[{\rm e}^{i\Lmat \Delta t/2}
\left(\Psi_n+{\Delta t\over 2} \Phi_1\right),t_{n+1/2}\right] \nonumber\\
\Phi_3&=& \Nmat\left[{\rm e}^{i\Lmat \Delta
t/2}\Psi_n+{\Delta t\over 2} \Phi_2,t_{n+1/2} \right] \nonumber\\
\Phi_4&=& \Nmat\left[{\rm e}^{i\Lmat \Delta
t}\Psi_n+\Delta t \Phi_3, t_{n+1} \right], \nonumber\\
\end{eqnarray}
which leads to (see Appendix \ref{app:RK}) 
\begin{equation}
\Psi_{n+1}= {\rm e}^{i\Lmat \Delta t} \Psi_{n}+{\Delta t\over 6}\left(\
\Phi_1+\Phi_2+\Phi_3+\Phi_4\right).
\end{equation}

\subsection{Exponential time differencing with constant nonlinear term
(ETD1)}
Assuming that the nonlinear term is constant during the time step
$t_n\rightarrow t_{n+1}$, i.e. 
\begin{equation}
\Nmat(\Psi,\tau) =\Nmat(\Psi_n,t_n) \ \ \ \ \  (t_n<\tau <t_{n+1}),
\end{equation}
the time propagation in Eq. \eqref{etd} becomes
\begin{equation}
\Psi_{n+1}= {\rm e}^{i\Lmat \Delta t} \Psi_{n}+\mathbf{M}_1 \Nmat(\Psi_n,t_n),
\end{equation}
where
\begin{equation}
\mathbf{M}_1=\Lmat^{-1} \left( {\rm e}^{i\Lmat \Delta t}-\mathbf{I}\right).
\end{equation}
This version of ETD is used in computational electrodynamics
\cite{taf05} and is ${\cal O}(\Delta t^2)$ accurate in time.

\subsection{Exponential time differencing with linearly raising nonlinear term
(ETD2)}
A better approximation is to take the nonlinear term as
\begin{multline}
\Nmat(\Psi,\tau) = \\ \Nmat(\Psi_n,t_n) + {\Delta \Nmat\over \Delta t} (\tau-t_n) \ \ \ \ \  (t_n<\tau <t_{n+1}),
\end{multline}
where $\Delta \Nmat = \left[\Nmat(\Psi_{n},t_{n})-\Nmat(\Psi_{n-1},t_{n-1})\right]$.
Now the time propagation in Eq. \eqref{etd} becomes
\begin{equation}
\Psi_{n+1}= {\rm e}^{i\Lmat \Delta t} \Psi_{n}+\mathbf{M}_1 \Nmat(\Psi_{n},t_{n})-{i\over\Delta
t} \mathbf{M}_2\Delta \Nmat,
\end{equation}
where
\begin{equation}
\mathbf{M}_2=\Lmat^{-1} \left(\mathbf{M}_1+{i\Delta t}\mathbf{I}\right).
\end{equation}
The method is ${\cal O}(\Delta t^3)$ accurate.

\subsection{Exponential time differencing with constant nonlinear
term, separating the wave function
(ETDCN)}
Exploring another possibility by assuming that the nonlinear potential is constant during the time step 
$t_n\rightarrow t_{n+1}$, i.e.
\begin{equation}
\Vmat^\text{N}(\Psi,\tau) =\Vmat(\Psi_n,t_n) \ \ \ \ \  (t_n<\tau <t_{n+1}),
\end{equation}
one can integrate Eq. \eqref{etd} using the trapezoidal rule, and 
the time propagation becomes
\begin{multline}
\left[\mathbf{I}+i\Vmat(\Psi_n,t_n)\Delta t\right]\Psi_{n+1}= \\{\rm e}^{i\Lmat \Delta t}
\left[\mathbf{I}-i\Vmat(\Psi_n,t_n)\Delta t\right]\Psi_{n}.
\end{multline}
This is similar to Crank--Nicolson propagation but with an extra ${\rm e}^{i\Lmat \Delta t}$
factor. By evaluating the inverse of the leftmost operator, one may arrive at a time evolution 
method which approximates the discrete time step propagator,
 \begin{multline}
\Psi_{n+1}= \\\left[\mathbf{I}+i\Vmat(\Psi_n,t_n)\Delta t\right]^{-1}{\rm e}^{i\Lmat \Delta t}
\left[\mathbf{I}-i\Vmat(\Psi_n,t_n)\Delta t\right]\Psi_{n}.
\end{multline}

\subsection{Exponential time differencing with second-order Runge--Kutta
(ETDRK2)}
The next level of approximation is to use a RK2 time
step ETD by introducing
\begin{multline}
\Nmat(\Psi,\tau) =\\\Nmat(\Psi_{n},t_{n}) + {\Delta \Nmat_a\over \Delta t} (\tau-t_n) \ \ \ \ \  (t_n<\tau <t_{n+1}),
\end{multline}
where $\Delta \Nmat_a=\left[\Nmat(\Phi_a,t_{n})-\Nmat(\Psi_{n-1},t_{n-1})\right]$ and
\begin{equation}
\Phi_a= {\rm e}^{i\Lmat \Delta t} \Psi_{n}+\mathbf{M}_1 \Nmat(\Psi_{n},t_{n}).
\end{equation}
This results in
\begin{equation}
\Psi_{n+1}=\Phi_a-{i\over\Delta t} \mathbf{M}_2\Delta \Nmat_a.
\end{equation}
Note that in the context of TDDFT, $\Nmat(\Phi_a,t_n)$ means that the Hartree and exchange-correlation
potentials must be calculated using the density defined by $\Phi_a$. The
accuracy is ${\cal O}(\Delta t^3)$.

\subsection{Exponential time differencing with RK4
(ETDRK4)}
The most accurate and stable approach is considered to be the RK4 method which is ${\cal O}(\Delta t^3)$ accurate
\cite{COX2002430}. This approach is a bit more involved than
the previous ones. One must define the following vectors:
\begin{eqnarray}
\Phi_a&=& \varphi_0\left({h\over 2}\Lmat\right) \Psi_n+ {h\over 2}\varphi_1\left({h\over 2}\Lmat\right)
\Nmat(\Psi_n,t_n),\\
\Phi_b&=& \varphi_0\left({h\over 2}\Lmat\right) \Psi_n+ {h\over 2}\varphi_1\left({h\over 2}\Lmat\right) \Nmat(\Phi_a,t_{n+1/2}),\nonumber\\
\Phi_c&=& \varphi_0\left({h\over 2}\Lmat\right) \Phi_a+ {h\over 2}\varphi_1\left({h\over 2}\Lmat\right)
\left[2\Nmat(\Phi_b,t_{n+1/2})-\Nmat(\Psi_n,t_n)\right],\nonumber
\end{eqnarray}
where $h=-i\Delta t$ and the $\varphi$-functions are defined in
Appendix \ref{app:phi}. 
One can then construct 
\begin{widetext}
\begin{equation}
\Psi_{n+1}={\rm e}^{-i\Lmat \Delta t} \Psi_{n}+h\left[
\varphi_1(h\Lmat)\mathbf{K}_1 
+\varphi_2(h\Lmat)\mathbf{K}_2 
+\varphi_3(h\Lmat)\mathbf{K}_3 \right],
\end{equation}
where
\begin{align}
\mathbf{K}_1 =&\Nmat(\Psi_n,t_n) \notag \\
\mathbf{K}_2 =&-3 \Nmat(\Psi_n,t_n)+2 \Nmat(\Phi_a,t_{n+1/2})
+2\Nmat(\Phi_b,t_{n+1/2}) - \Nmat(\Phi_c,t_{n+1}) 
 \notag \\
\mathbf{K}_2
=&4\left[\Nmat(\Psi_n,t_n)-\Nmat(\Phi_a,t_{n+1/2})-\Nmat(\Phi_b,t_{n+1/2})
+ \Nmat(\Phi_c,t_{n+1}) \right].
\end{align}
\end{widetext}
In this case the small eigenvalues lead  to numerical problems in
the calculation of the inverse matrix, and, following Ref. \cite{COX2002430},
we simply eliminate these small eigenvectors from inverse.

\subsection{Krogstad time propagation (Krogstad)}
The ETDRK4 approach was further developed in Ref. \cite{Krogstad:2005:GIF:1063068.1063072}
using a truncated Taylor expansion of the nonlinear part in order to increase
the accuracy.  It only differs from the  ETDRK4 method in the definition of the $\Phi_a$, $\Phi_b$, and $\Phi_c$ functions:
\begin{widetext}
\begin{eqnarray}
\Phi_a&=& \varphi_0\left({h\over 2}\Lmat\right) \Psi_n+ {h\over 2}\varphi_1\left({h\over 2}\Lmat\right)
\Nmat(\Psi_n,t_n),\\
\Phi_b&=& \varphi_0\left({h\over 2}\Lmat\right) \Psi_n+ {h\over 2}\varphi_1\left({h\over 2}\Lmat\right) \Nmat(\Psi_n,t_n) + h\varphi_2\left({h\over 2}\Lmat\right) \left[\Nmat(\Phi_a,t_{n+1/2}) - \Nmat(\Psi_n,t_n)\right],\nonumber\\
\Phi_c&=& \varphi_0\left({h}\Lmat\right)\Psi_n + h\varphi_1\left(h\Lmat\right)\Nmat(\Psi_n,t_n) + 2h \varphi_2\left(h\Lmat\right)\left[\Nmat(\Phi_b,t_{n+1/2})-\Nmat(\Psi_n,t_n)\right]
.\nonumber
\end{eqnarray}
\end{widetext}

\section{Numerical results}\label{sec:results}
To test these approaches we use a simple one-dimensional helium atom model that
has been often used in similar test calculations
\cite{PhysRevA.95.013414}. The Hamiltonian in atomic units is 
\begin{equation}
H=-{1\over 2} {d^2\over dx^2}+V(x)+E(t)x+V_H[\rho(x,t)]+V_{ex}[\rho(x,t)].
\label{ham}
\end{equation}
In this equation, $V(x)$ is a soft Coulomb potential given
as \cite{su1991model,Castro2015}
\begin{equation}
V(x)=-\dfrac{2}{(a^2 + x^2)^{1/2}},
\end{equation}
where $a$ has been set to unity. The two-electron density is defined as either
\begin{equation}
\rho(x,t)=2\lvert\Psi_1(x,t)\rvert^2 \ \ \  \rm (model\  A)
\end{equation}
or
\begin{equation}\label{eq:modelB}
\rho(x,t)=\lvert\Psi_1(x,t)\rvert^2+ \lvert\Psi_2(x,t)\rvert^2  \ \ \  \rm
(model\  B),
\end{equation}
where $\Psi_1$ and $\Psi_2$ are the ground and first excited states, 
respectively, of the Hamiltonian at $t=0$. Model A is an uncoupled 
system, while, in model B, the two
states are coupled, leading to more complicated nonlinear effects. 
The Hartree potential is calculated as 
\begin{equation}
V_H(x)=\int dy {\rho(y)\over \sqrt{(x-y)^2+a^2}},
\end{equation}
where the potential is softened using the same value for $a$.
The exchange-correlation potential is given by the exact-exchange
approximation \cite{Castro2015}
\begin{equation}
V_{ex}(x)=-{1\over 2} V_H(x).
\end{equation}

The time-dependent term $E(t)x$ corresponds to the contribution of the
electric field, used to represent a laser pulse,
under the dipole approximation. This term may be incorporated within either
the linear part---time-dependent $\Lmat$---or the nonlinear part---time-independent
$\Lmat$. In the case of the latter, the two parts take the form  
\begin{equation}\label{eq:L_no_laser}
L=-{1\over 2} {d^2\over dx^2}+V(x) 
\end{equation}
\begin{equation}\label{eq:N_laser}
\Nmat(\Psi,t)=\left(E(t)x+V_H[\rho(x,t)]+V_{ex}[\rho(x,t)]\right) \Psi(x,t).
\end{equation}

Two different types of TDDFT calculations were performed using our model one-dimensional helium system.
In the first, the system was placed in an excited state at the beginning of the simulation.
Such an excited state causes fluctuations in the electron density,
and this change in density, in turn, causes the nonlinear potential to change rapidly.
In these calculations, no additional time-dependent potential was added, resulting in 
only the nonlinear potential being time-dependent.
These simulations were carried out within a computational box of width 160 Bohr, and
a complex absorbing potential (CAP) \cite{doi:10.1063/1.1517042} was added, so as to allow some ionization which occurs early in the simulation.
In simulations of the second type, the electrons were subject to a time-dependent laser potential,
represented using the dipole approximation, $V_\text{laser}=E(t)x$, where the form of the 
electric field was chosen as a variation of the smooth turn-on pulse
\cite{compnano_varga},
\begin{equation}
\label{eq:laser}
E(t)=
\begin{cases}
\hphantom{-}E_0\sin\left(\dfrac{\pi t}{2T_c}\right)\sin(\omega t), &\text{if $0
\leq t \leq T_c$},\\[2ex]
\hphantom{-}{E_0}\sin(\omega t), &\text{otherwise}.
\end{cases}
\end{equation}
In these calculations, the parameters $\omega$ and
$T_c$ were set to 0.148 and $6/\omega$, respectively. Simulations were carried out for  
maximum electric fields, $E_0$, of both 0.1 a.u. and 1.0 a.u. and a computational box of width 400 Bohr.
The initial states, $\Psi(x,t=0)$, were calculated by diagonalizing the
Hamiltonian, Eq. \eqref{ham}, with $E(t=0)=0$. The Hamiltonian was
represented using a pseudospectral basis \cite{compnano_varga}.

A benchmark calculation was performed using the Taylor time propagator with a time step of 0.00001 a.u. (1000 times smaller than necessary for stability). The wave function from various methods is compared to the benchmark at various times using the Tanimoto index \cite{QUA:QUA560420607}
\begin{equation}
\label{eq:psi_metric1}
\sigma_i(t) = {I_{Bi}(t) \over I_{BB}(t) + I_{ii}(t) - I_{Bi}(t)},
\end{equation}
where 
\begin{equation}
\label{eq:psi_metric1_int}
I_{ij}(t) = \int  \lvert\Psi_i^*(t) \Psi_j(t) \rvert dx,
\end{equation}
with $B$ indicating the wave function from the benchmark calculation.
This metric ranges in value from 0 to 1, with 1 indicating a perfect match.
The wave functions are not normalized prior to comparison
since this similarity method will take into account whether or not the functions differ by a constant. 
The time-averaged agreement between the benchmark and propagated wave function is tracked by $\sigma_T$.
In practice, we determine the time-averaged agreement within a given range, $T_i$ to $T_f$,
\begin{equation}
\label{eq:psi_metric_time}
\sigma_{(T_i\rightarrow T_f),i} = \int_{T_i}^{T_f}  \sigma_i(t) dt
\end{equation}
The time-averaged error of $\Psi$ is then taken to be $1-\sigma_T$.

\subsection{Excited state superposition}

\begin{figure}[h]
\centering
\includegraphics[width=\linewidth]{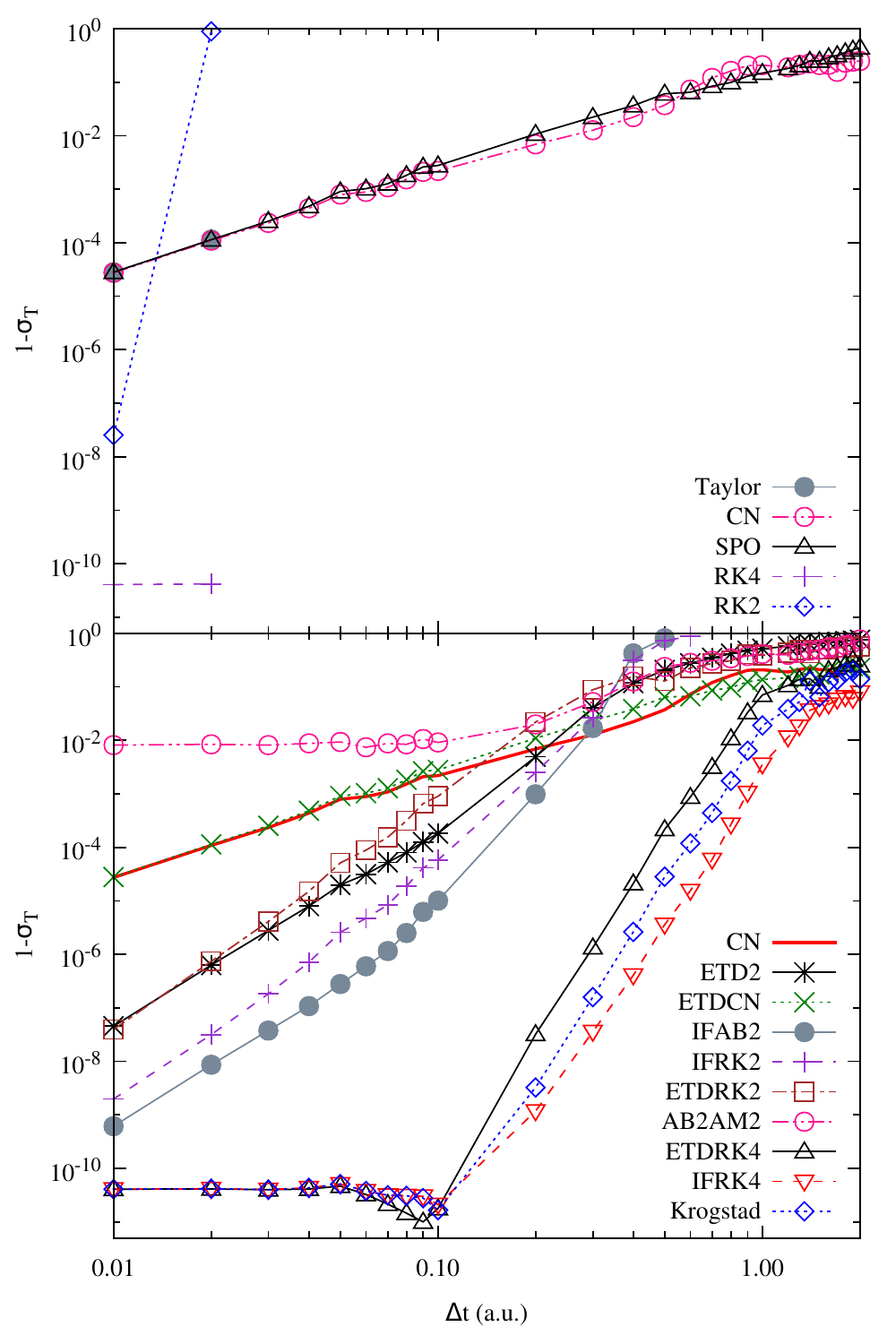}
\caption{The time-averaged error of various methods for integrating the TDKS equations when electrons are initialized in an excited state.
The time interval considered was between $T_i=10$ a.u. and $T_f=100$ a.u. Time evolution 
operator approaches are shown above while IMEX, IF, and ETD approaches are shown below. CN is shown in the latter for comparison. 
} 
\label{fig:excitedshort}
\end{figure}

We first consider a one-orbital case, defined 
as an equal combination of the ground and first excited states at $t=0$.
The system, upon such initialization, is free to develop without external perturbation.
The nonlinear term is, therefore, the only part that is time-dependent.
The performance of various methods for integrating the TDKS equations is shown in Fig. \ref{fig:excitedshort} for a simulation up to $T_f$ = 100 a.u.
Considering the Taylor, CN, and
SPO methods from this set of calculations,
the error in each appears comparable for all that are stable.
The Taylor time propagation yields results comparable to CN and split operator for time step sizes up to dt=0.02 a.u., after which it fails.
To maintain $99\%$ accuracy of the wave function, the largest time step size possible for the time evolution operator methods is determined to be around 0.2 a.u. 
As for the the simple Runge--Kutta methods (see Fig.
\ref{fig:excitedshort}), these
have less error than the time evolution operator methods; they are
limited by the maximum time step size.

\begin{figure}[h]
\centering
\includegraphics[width=\linewidth]{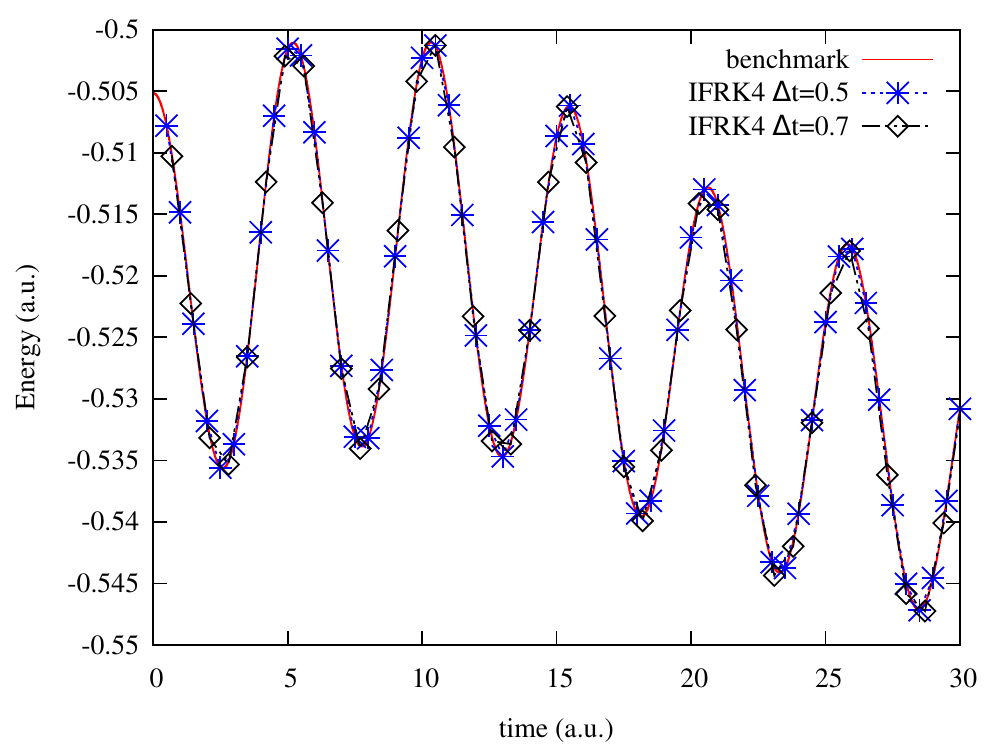}
\caption{The time-dependent energy of the IFRK4 integration method using time step sizes of 0.5 and 0.7 a.u.  as compared to a benchmark calculation.
} 
\label{fig:excitedene}
\end{figure}

When comparing the time evolution operator and simple Runge--Kutta methods to the IF and ETD methods,
the latter are seen to perform much better, with less error and larger allowed time step sizes.
Of methods that use RK2-type time integration or are ${\cal O}(\Delta t^3)$ accurate, the IFRK2 and IFAB2 methods perform marginally better than the ETDRK2 method. For each, the maximum time step size maintaining $99\%$ accuracy is around 0.2 to 0.3 a.u. 
Of methods that use RK4-type time integration or are ${\cal O}(\Delta
t^4)$ accurate, IFRK4 does best, with Krogstad outperforming ETDRK4 integration and a maximum time step size for each methods near 1.0 a.u. While the norm of the wave function is not conserved for all methods up to this maximum time step size, a CAP was used in these calculations; therefore, the norm is expected to deviate from unity, regardless.
The energy oscillates quickly with a period of 5 a.u., shown in Fig. \ref{fig:excitedene}. Nevertheless, the IFRK4 method is capable of accurately producing the proper energy curve for even large time step sizes of 0.7 a.u.   

\begin{figure}[h]
\centering
\includegraphics[width=\linewidth]{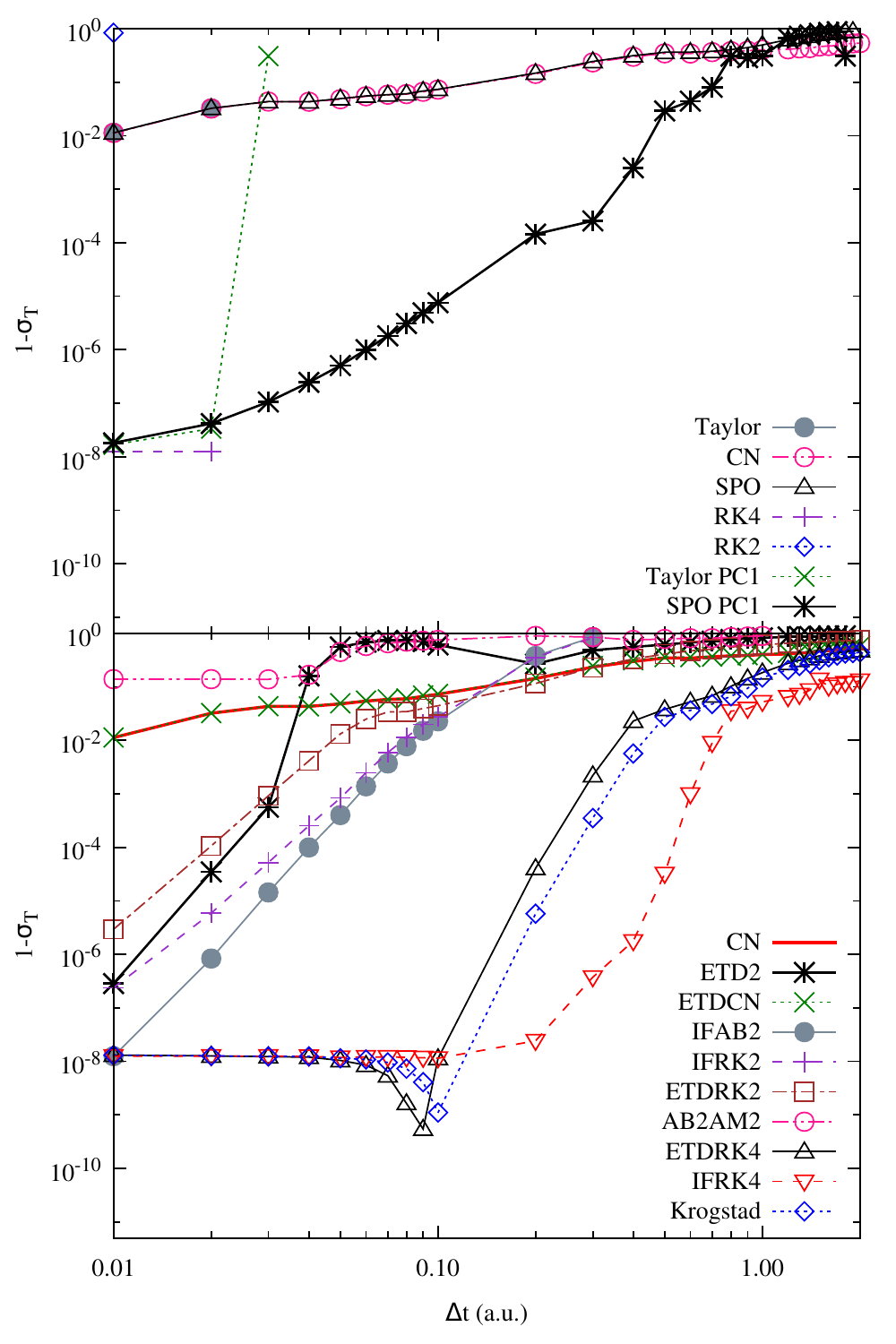}
\caption{The time-averaged error of various methods for integrating the TDKS equation over a long time period.
The time interval considered was between $T_i=10$ a.u. and $T_f=1000$ a.u. Time evolution 
operator approaches are shown above while IMEX, IF, and ETD approaches are shown below. CN is shown in the latter for comparison.
} 
\label{fig:excitedlong}
\end{figure}

For longer simulations, the nonlinear nature of the TDKS equations requires accurate integration in time. As a rigorous test of the methods considered in this work, a long ($T_f=1000$ a.u.) simulation was performed using the same initial state as described above. A comparison of all methods for the longer simulation is shown in Fig. \ref{fig:excitedlong}. The trends are similar to the shorter simulation, though some of the methods are more effected by the error accrued in time. 

\begin{figure}[h]
\centering
\includegraphics[width=\linewidth]{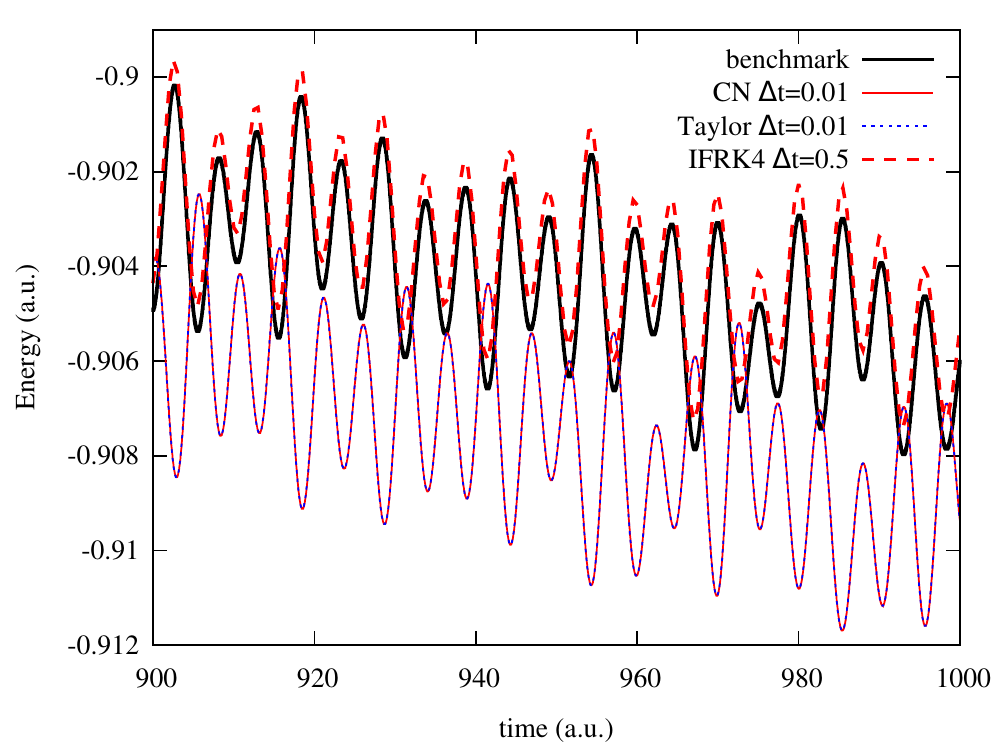}
\caption{The energy as a function of time for various methods. The time evolution operator methods (Taylor and CN) drift slowly away from the exact solution, however they overlap one another. The divergence of such methods from the converged solution is more noticeable for longer simulations.
} 
\label{fig:excitedlong_ene}
\end{figure}

For the longer time trials, the time evolution operator approaches fail to correctly integrate the wave function over the course of the simulation. This is because the Hamiltonian is assumed to be constant between time steps and the error associated with this approximation accrues throughout the simulation and is exacerbated by the nonlinear potential. The time evolution operator methods can be improved by using predictor--corrector schemes; Taylor and SPO with predictor--corrector perform similarly to RK4, see Fig. \ref{fig:excitedlong}. The energy as a function of time is shown in Fig. \ref{fig:excitedlong_ene} for several methods towards the end of the simulation. The energy difference is quite noticeable in both CN and Taylor methods, though they match well with one another. This result indicates that even trusted methods, like CN, may fail for longer simulation times due to the inclusion of a nonlinear potential. 
For the IF and ETD methods, the longer simulation shows that RK2-type methods do not perform as well as their RK4 counterparts. The maximum time step sizes for the ETDRK2, IFRK2, and IFAB2 methods are near 0.08 a.u. For the ETDRK4 and Krogstad methods, the maximum time step size is near 0.3 a.u., while IFRK4 seems to perform well until 0.5 a.u.

\subsection{Laser with one orbital}
The collection of methods was also tested with applications to laser-driven dynamics. In these simulations, the wave function was initialized in the ground state, with the density defined using model A, and a time-dependent electric field of the form shown in Eq. \eqref{eq:laser} was applied. The associated, additional term in the Hamiltonian, $V_\text{laser}=E(t)x$, was first included in the nonlinear part such that $\Lmat$ and $\Nmat$ were of the forms given in Eqs. \eqref{eq:L_no_laser} and \eqref{eq:N_laser}. Secondly, this linear term was added to the linear part for comparison of accuracies. In the latter scenario, the matrix-valued functions containing $\Lmat$ were updated at each time step. While, in principle, the external electric field would be expected to cause ionization, a large simulation box length of 400 Bohr was used, so no CAP was deemed necessary. 

\begin{figure}[h]
\centering
\includegraphics[width=\linewidth]{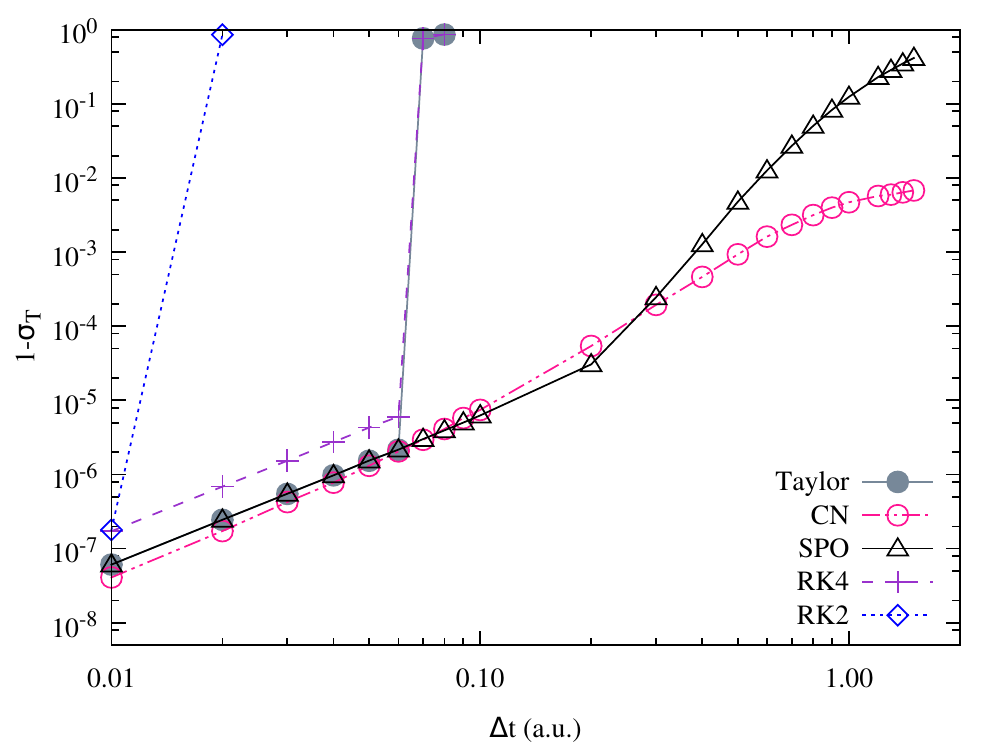}
\caption{The time-averaged error of various methods for integrating the TDKS equation when electrons are driven by an external electric field of strength 0.1 a.u. The time-dependent potential from the electric field was included in the linear part.
The time interval considered was between $T_i=10$ a.u. and $T_f=100$ a.u.
} 
\label{fig:laser_time_evo}
\end{figure}

The accuracy of the various time evolution methods is shown in Fig. \ref{fig:laser_time_evo}. These methods performed much better here than in the case of the excited state superposition. It is likely because the orbitals develop more slowly under the influence of the ramped laser, which drives their dynamics in this case, rather than the nonlinear part. This allows the approximation of a constant Hamiltonian at each time step to better describe the dynamics. 

\subsubsection{Laser potential in nonlinear part} 

\begin{figure}[h]
\centering
\includegraphics[width=\linewidth]{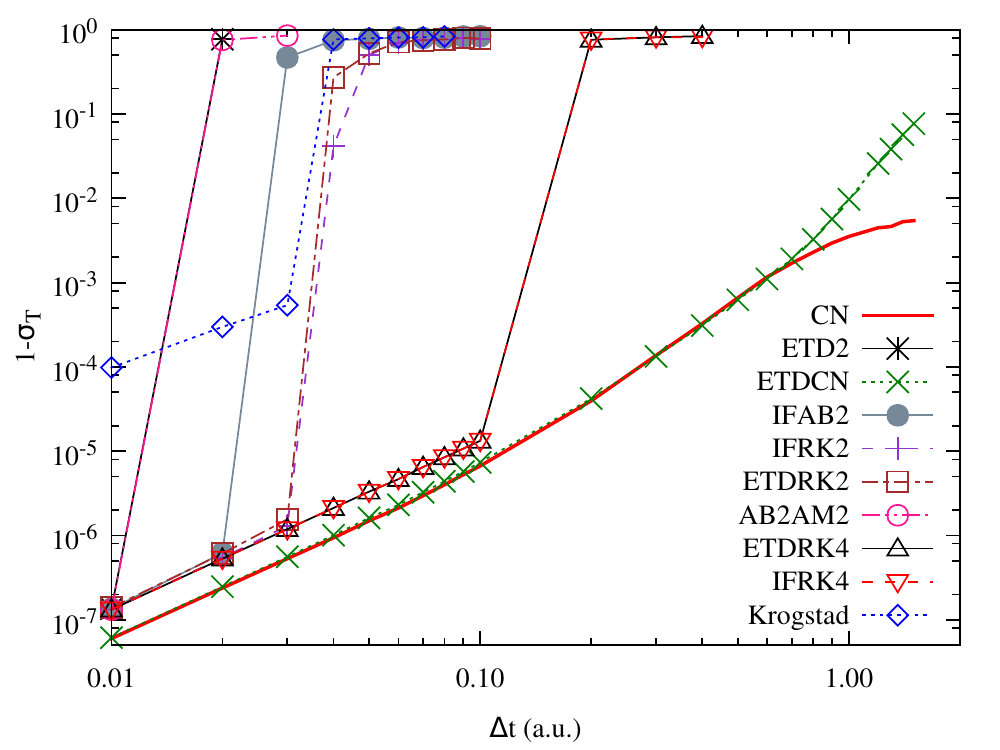}
\caption{The time-averaged error of various methods for integrating the TDKS equation when electrons are driven by an external electric field of strength 0.1 a.u. The time-dependent potential from the electric field was included in the nonlinear part. The time interval considered was between $T_i=10$ a.u. and $T_i=100$ a.u.
CN is shown for comparison.
} 
\label{fig:laser_err_1oef0p1_lWnl}
\end{figure}

By including the potential from the laser in the nonlinear part, the linear part remains time-independent, and matrix-valued functions containing $\Lmat$ may be precalculated before time propagation and used throughout.
The errors related to the various IMEX, IF, and ETD methods for an external field strength of 0.1 a.u. are shown in Fig. \ref{fig:laser_err_1oef0p1_lWnl}.
Methods using RK4-type integrations, other than Krogstad, exhibit stability for time step sizes up to 0.1 a.u., whereas those using RK2-type integration remain stable only for time step sizes below 0.03 a.u. For stable time step sizes, these methods yield accuracies within an order of magnitude of the CN method. However, it appears that the separate numerical integration of the time-dependent nonlinear part hinders the IF and ETD methods such that they are outperformed by the CN method for all choices of time step size. ETDCN is able to match the accuracy of CN for time step sizes up to 0.7 a.u. due to it's time evolution form.

\subsubsection{Laser potential in linear part} 

\begin{figure}[h]
\centering
\includegraphics[width=\linewidth]{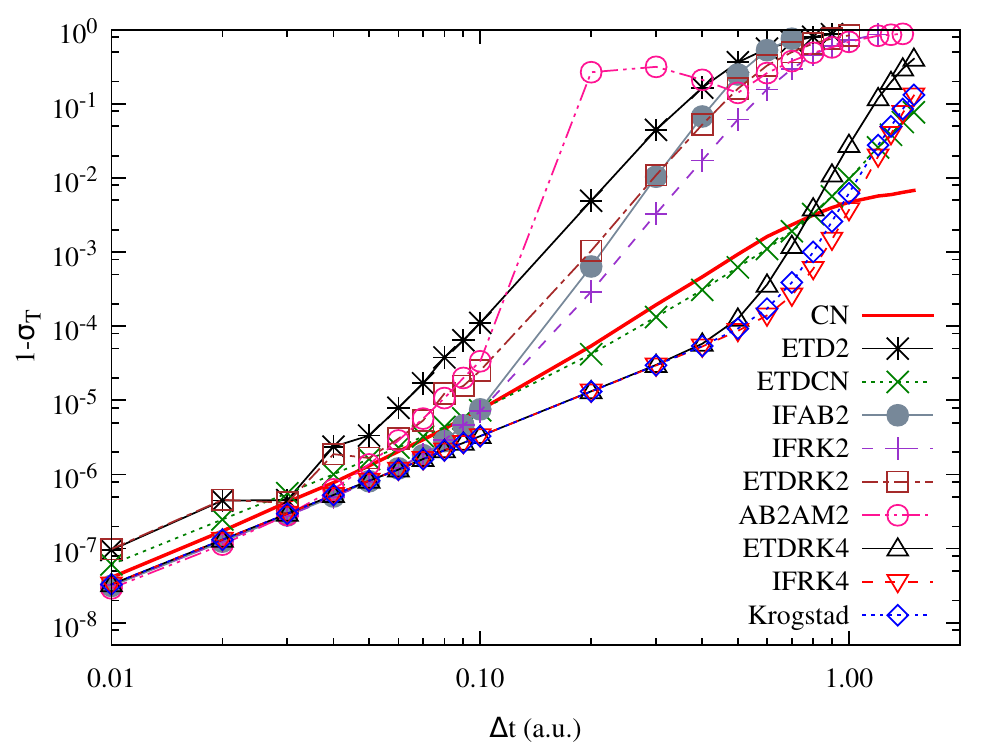}
\caption{The time-averaged error of various methods for integrating the TDKS equation when electrons are driven by an external electric field of strength 0.1 a.u. The time-dependent potential from the electric field was included in the linear part.
The time interval considered was between $T_i=10$ a.u. and $T_f=100$ a.u. CN is shown for comparison.
} 
\label{fig:laser_err_1oef0p1}
\end{figure}

By including the 0.1 a.u. field strength laser in the linear part, the stability of the IF and ETD methods is significantly improved, as shown in Fig. \ref{fig:laser_err_1oef0p1}. Here, there is a clear grouping of ${\cal O}(\Delta t^3)$ and ${\cal O}(\Delta t^4)$ methods. 
RK4-type methods are shown to outperform CN for choices of time step size up to about 0.8 a.u.

\begin{figure}[h]
\centering
\includegraphics[width=\linewidth]{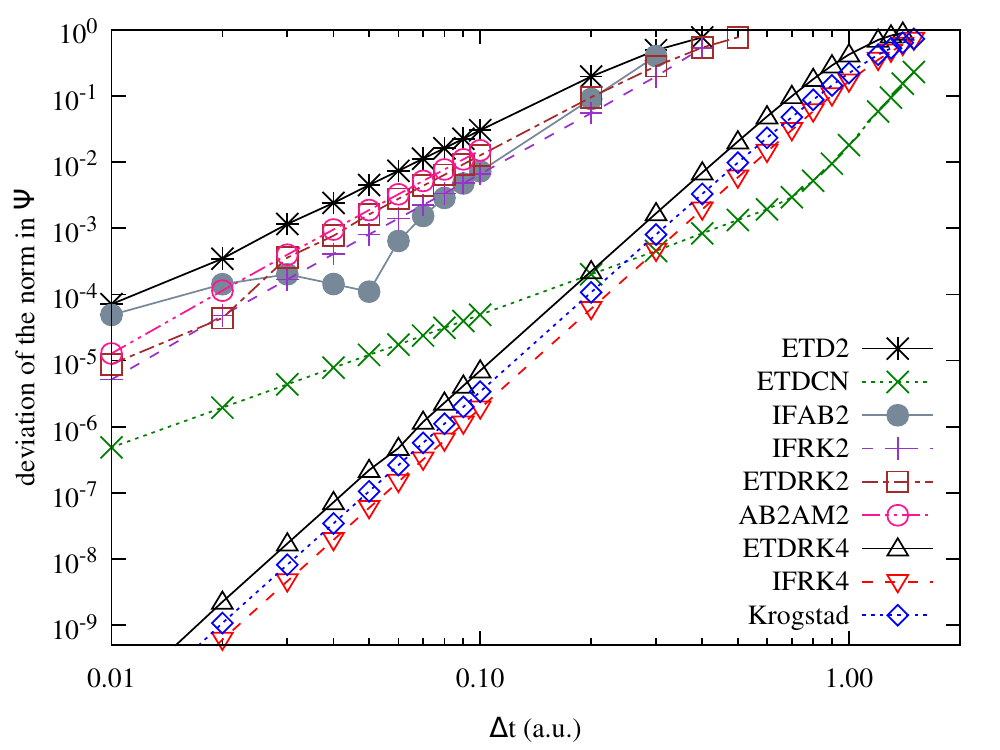}
\caption{The difference of the norm from unity for various methods when electrons are driven by an external electric field of strength 0.1 a.u. The time-dependent potential from the electric field was included in the linear part.
The time interval considered was between $T_i=10$ a.u. and $T_f=100$ a.u.
} 
\label{fig:laser_norm_1oef0p1}
\end{figure}

Ideally, the norm of the wave function should remain at unity; however, for some methods, this value deviates.
The norm conservations of the IMEX, IF, and ETD methods are shown in Fig. \ref{fig:laser_norm_1oef0p1}. Here, the RK4-type methods perform best for time step sizes below about 0.2 a.u., while ETDCN does so otherwise. Overall, IFRK4 maintains a slight advantage over the other RK4 exponential integrator methods. In principle, for cases such as this one where a CAP is not necessary, one may renormalize the orbitals at each time step, eliminating this divergence as a source of error. 

\begin{figure}[h]
\centering
\includegraphics[width=\linewidth]{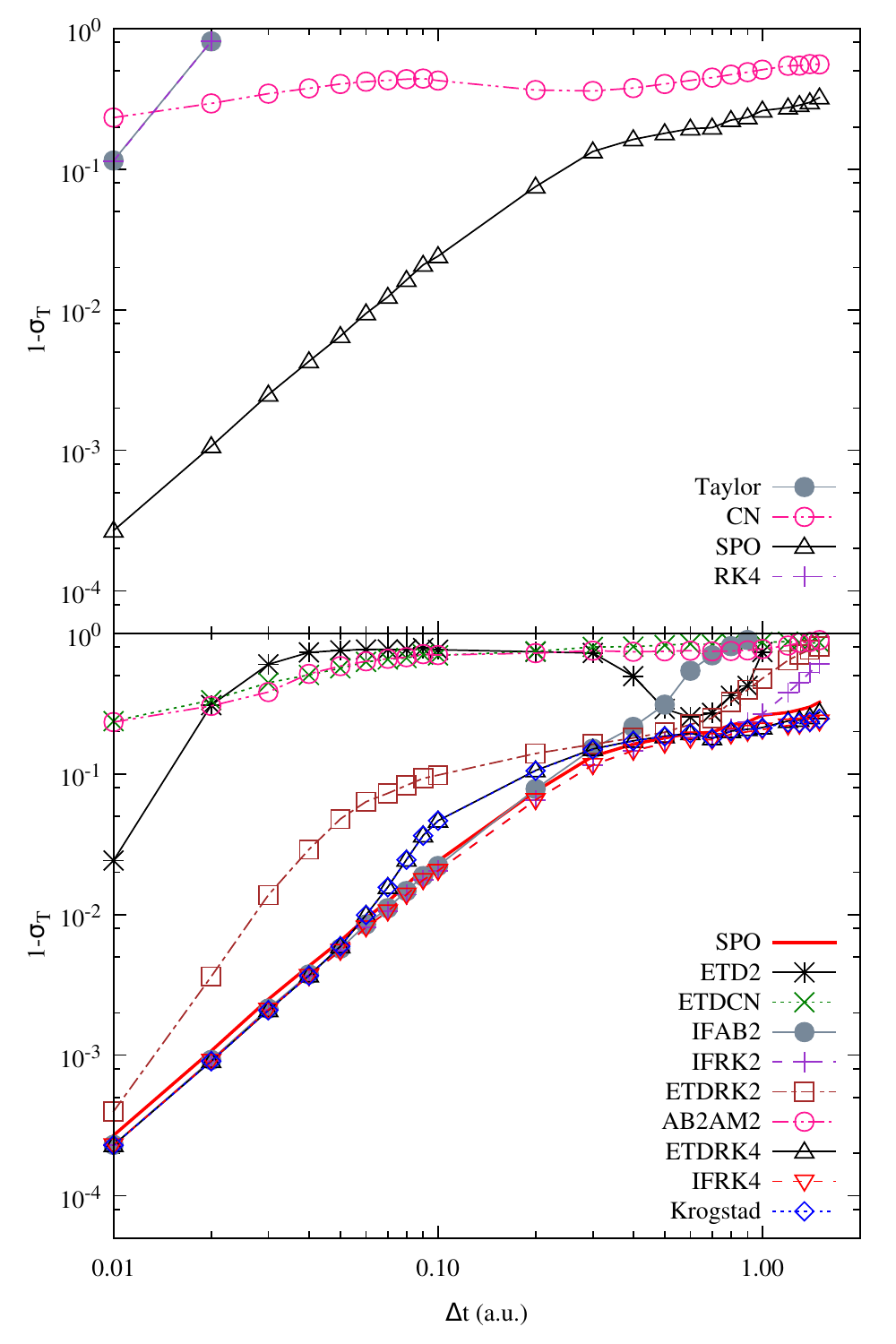}
\caption{The time-averaged error of various methods for integrating the TDKS equation when electrons are driven by a strong external electric field of strength 1.0 a.u.
The time-dependent potential from the electric field was included in the linear part. The time interval considered was between $T_i=10$ a.u. and $T_f=100$ a.u. SPO is shown for comparison. The second-order Runge--Kutta method is excluded from the top figure due to it being unstable for each choice of time step size.
} 
\label{fig:laser_err_1oef1}
\end{figure}

When a stronger laser field is considered, all methods generally perform worse, as presented in Fig. \ref{fig:laser_err_1oef1}. For the time evolution operator methods, this is due to the large magnitudes of the rapidly changing Hamiltonian which break down approximations of the exponential time propagator. Amongst these, the SPO approach performs best due to its analytic expression of the matrix exponential form. We note that these calculations were performed using the length gauge convention for the laser potential. By using velocity gauge with the SPO, one may better represent dynamics induced by high intensity lasers due to the method's equivalent formalism to propagation using a basis defined by free electrons reacting to an external field---the Volkov basis \cite{PhysRevA.95.013414}. In the case of the IF and ETD methods, this degradation in accuracy is due to the breakdown of the approximation that the $\Lmat$ matrix and its related matrix-valued functions are constant for the duration of any given time step. While SPO performs best of the time evolution methods, IFRK4 is able to match or beat it for all choices of time step size, while other RK4-type methods maintain similar accuracy. Notably, ETDRK2 and IFRK2 do very well in this case, with the latter being nearly indistinguishable from its RK4-type counterpart.  

\subsection{Coupled system -- Laser with two orbitals}
\begin{figure}[h]
\centering
\includegraphics[width=\linewidth]{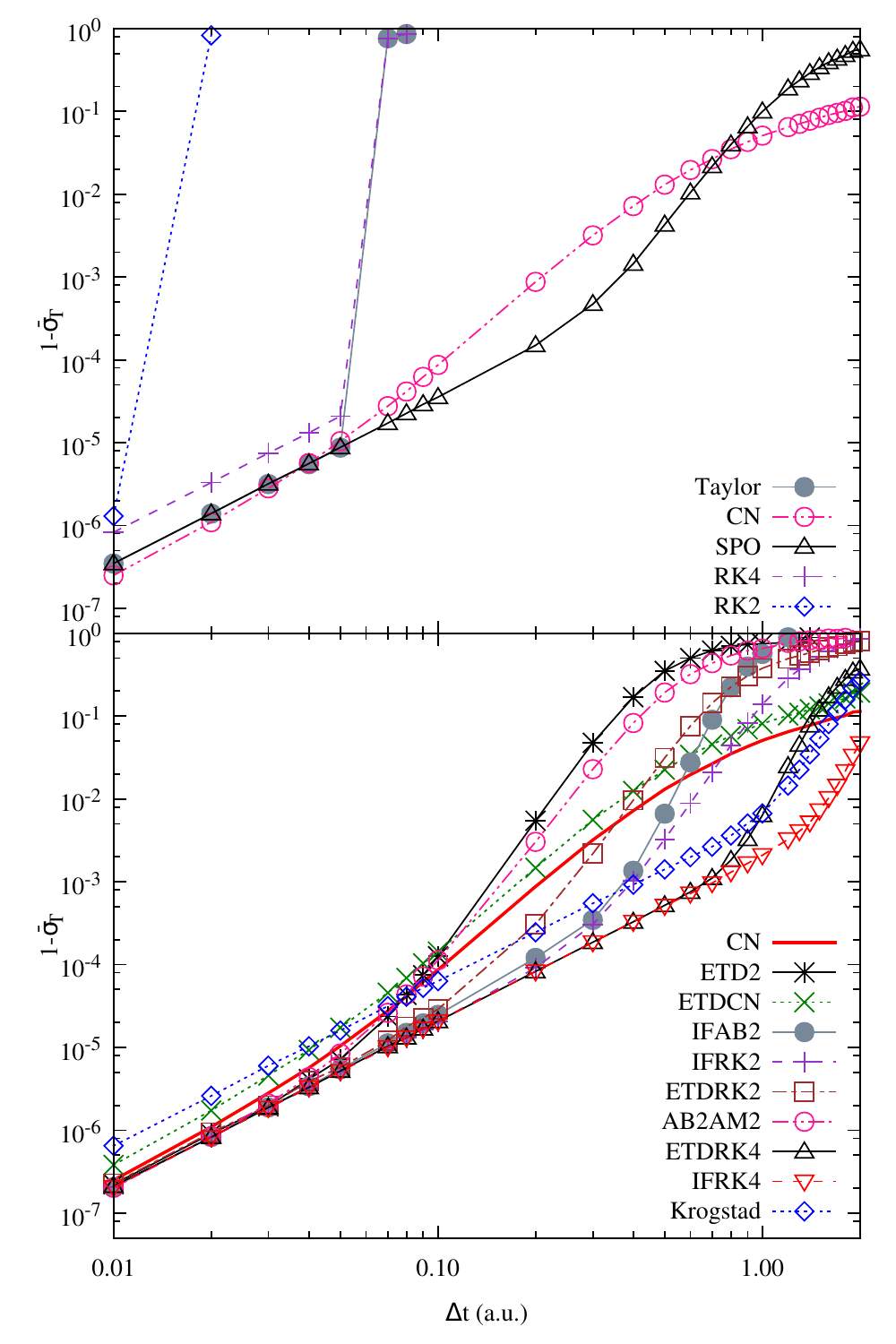}
\caption{The time- and orbital-averaged error for model B using various methods for integrating the TDKS equation when electrons are driven by an external electric field of strength 0.1 a.u. The time interval considered was between $T_i=10$ a.u. and $T_f=100$ a.u. Time evolution 
operator approaches are shown above while IMEX, IF, and ETD approaches are shown below. CN is shown in the latter for comparison.
} 
\label{fig:laser_err_2oef0p1}
\end{figure}

\begin{figure}[h]
\centering
\includegraphics[width=\linewidth]{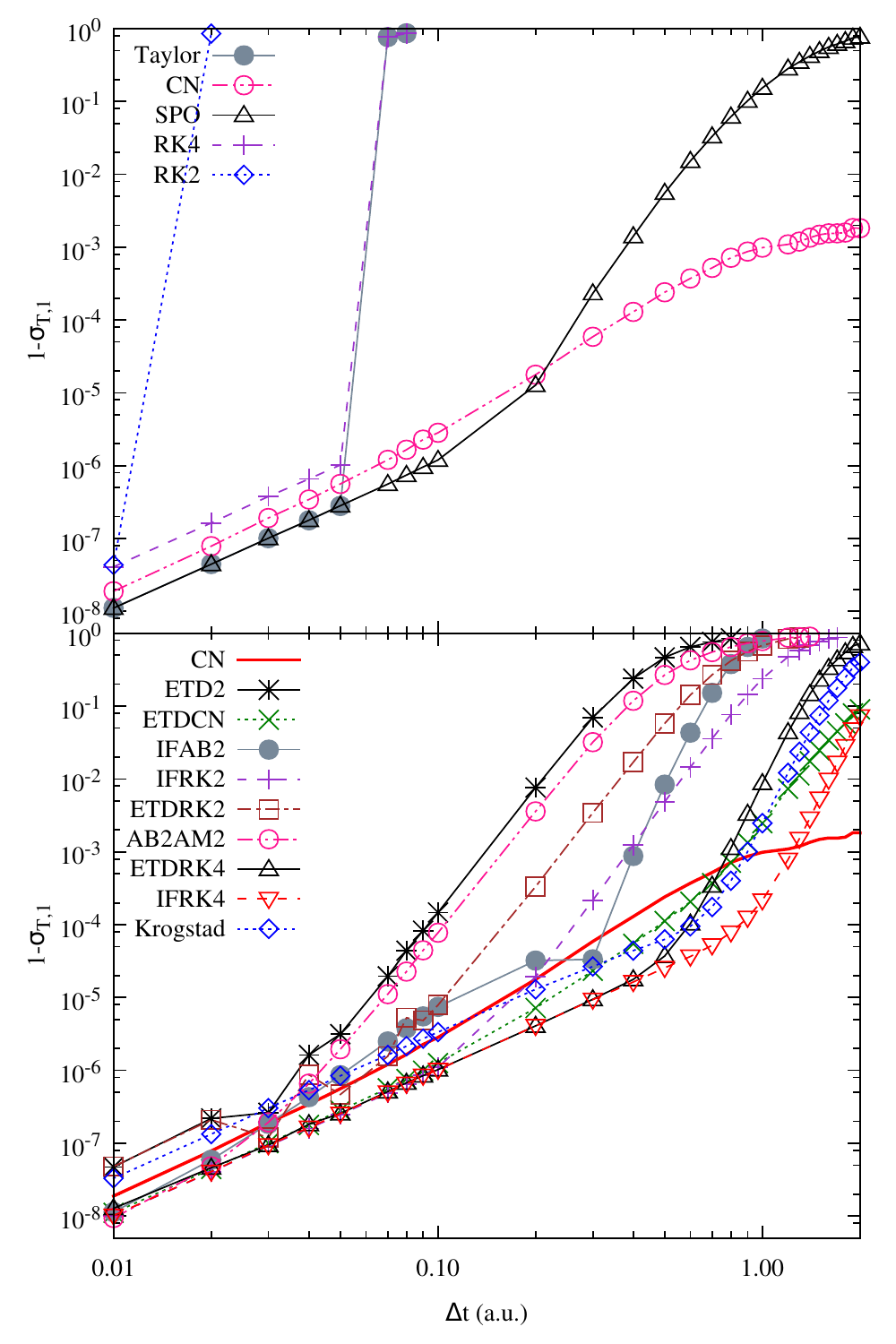}
\caption{The time-averaged error of orbital 1 for model B, using various methods for integrating the TDKS equation when electrons are driven by an external electric field of strength 0.1 a.u. The time interval considered was between $T_i=10$ a.u. and $T_f=100$ a.u. Time evolution 
operator approaches are shown above while IMEX, IF, and ETD approaches are shown below. CN is shown in the latter for comparison. 
} 
\label{fig:laser_err_2oef0p1_o1}
\end{figure}

\begin{figure}[h]
\centering
\includegraphics[width=\linewidth]{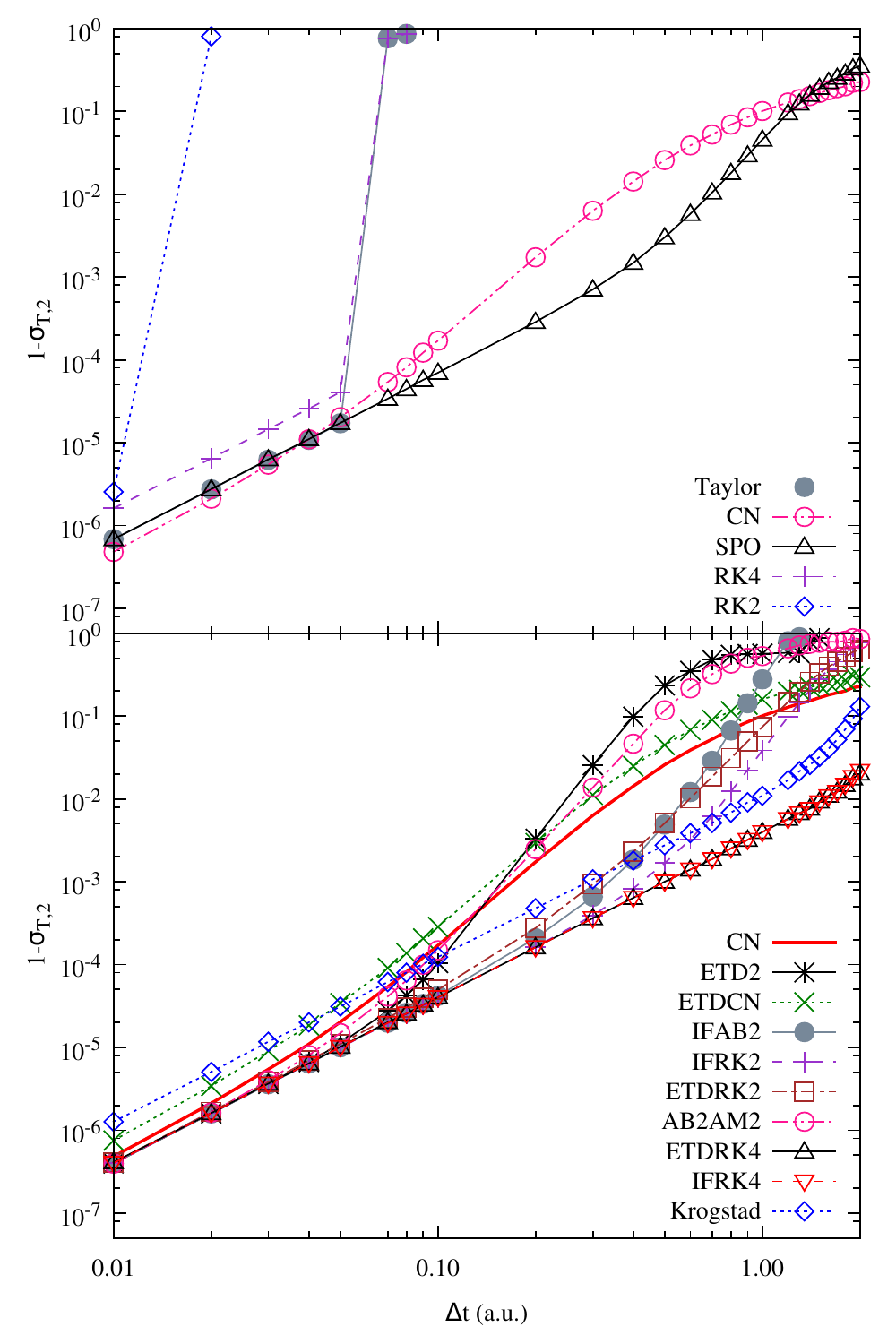}
\caption{The time-averaged error of orbital 2 for model B, using various methods for integrating the TDKS equation when electrons are driven by an external electric field of strength 0.1 a.u. The time interval considered was between $T_i=10$ a.u. and $T_f=100$ a.u. Time evolution 
operator approaches are shown above while IMEX, IF, and ETD approaches are shown below. CN is shown in the latter for comparison.
} 
\label{fig:laser_err_2oef0p1_o2}
\end{figure}

As a more rigorous test of the nonlinear contribution, a system comprised of two electrons in separate orbitals, coupled via the Hartree and exchange-correlation potentials, was time propagated under the influence of an external electric field using the collection of methods. This choice corresponds to model B, introduced in Eq. \eqref{eq:modelB}. The laser potential is included in the linear part, $\Lmat$, in the following results. Here, the similarity, $\sigma$, is taken as the average of the two orbitals, denoted as $\overline{\sigma}$. This average error is presented for the case of a peak electric field of 0.1 a.u. in Fig. \ref{fig:laser_err_2oef0p1}.
We can also compare the time-averaged error associated with each orbital separately to their respective benchmark, see Fig. \ref{fig:laser_err_2oef0p1_o1} and Fig. \ref{fig:laser_err_2oef0p1_o2}.

 We find that the second, higher energy orbital dominates as the larger source of error for most methods, which is to be expected due to it's spatial extension and more complicated nodal structure. The CN and SPO methods each take turns performing best of the time evolution methods for different choices of time step size and orbital. Of the IF and ETD methods, IFRK4 and ETDRK4 similarly exhibit the most accurate results for the second orbital, but the former gains an advantage in its representation of the first orbital. The orbital-averaged error indicates IFRK4 as the clear front-runner. All methods which remain stable behave similarly in the range of time step sizes below about 0.1 a.u.; however, above this point, methods of ${\cal O}(\Delta t^3)$ begin to accumulate larger amounts of error. Of these, IFRK2 performs best, with an time- and orbital-averaged error similar to the RK4-type methods up to time step sizes of about 0.3 a.u.
The reason the integration methods perform better, in general, for model B, is that the dynamics of the nonlinear potential change more slowly than for model A.

\begin{figure}[h]
\centering
\includegraphics[width=\linewidth]{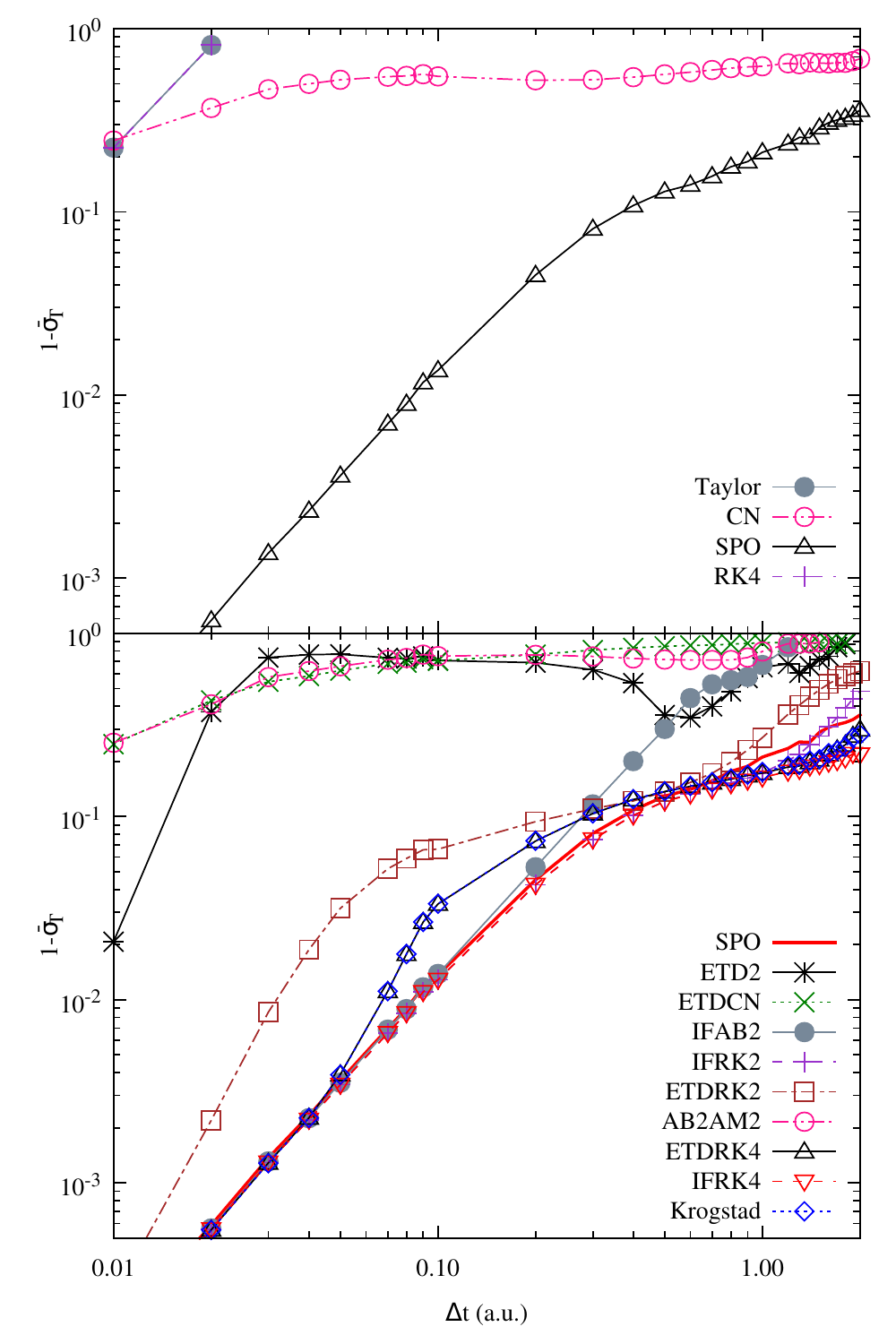}
\caption{The time- and orbital-averaged error for model B using various methods for integrating the TDKS equation when electrons are driven by an external electric field of strength 1.0 a.u. The time interval considered was between $T_i=10$ a.u. and $T_f=100$ a.u. Time evolution 
operator approaches are shown above while IMEX, IF, and ETD approaches are shown below. SPO is shown in the latter for comparison. The second-order Runge--Kutta method is excluded from the top figure due to it being unstable for each choice of time step size.  
} 
\label{fig:laser_err_2oef1p0strong}
\end{figure}

Similar to model A, for the case of an intense laser, the general accuracy of each method is lessened, as shown in Fig. \ref{fig:laser_err_2oef1p0strong}. Again, SPO performs best of the time evolution methods, IFRK4 remains most accurate, and IFRK2 outperforms or performs similarly to the RK4-type methods for any choice of time step size. 

\section{Summary}

We have implemented forms of IMEX and exponential integrator methods within TDDFT calculations and have compared the results to those of conventional time evolution methods. The cases studied included dynamics driven primarily by the nonlinear part of the Hamiltonian, as well as those driven primarily by a time-dependent, linear laser potential. We have found that of the time evolution methods, the CN and SPO methods performed best for the various test simulations. Typically these two approaches yielded similar results except in the case of intense laser fields, in which the SPO showed a definitive advantage. Of the IMEX and exponential integrator methods, the RK4-type IF and ETD approaches yielded the most accurate results for each of the test cases.

Comparing the leading methods of both groups, the RK4-type exponential integrator methods were able to match or exceed the accuracy of the leading time evolution methods in each set of tests. For dynamics driven by a linear, time-dependent potential, the RK4-type exponential integrator methods were able to match the front-runners of the time evolution group,  CN or SPO, for both moderate and high laser intensities. In cases where the dynamics were driven by the nonlinear part of the Hamiltonian, the RK4-type exponential integrator methods outperformed even the best suited time evolution methods by orders of magnitude.          


While the ETD method is typically seen as being the most accurate of
the exponential integrators in the mathematical literature, in our
results, the IF method performed uniformly better, though slightly so. 
This may be due to a more complicated structure of the nonlinear
part in Eq. \eqref{etd} for TDDFT rather than in other equations investigated
in the literature where the nonlinear part is typically a $y^k$ term
(in Eq. \eqref{main}). 

Beyond the success of the RK4-type exponential integrators shown in this study, 
we note that they may further benefit from the ability of Runge--Kutta 
approaches to propagate the wave function using variable time step sizes. 
This implies the capability of dynamically adjusting the time step size 
throughout simulations in order to best balance the computational cost and accuracy. 
While the time step size has been kept fixed for each calculation presented 
in this work, we propose such a modification as a subject for future improvement. 

In tests including a time-dependent, linear potential associated with a driving laser field, the accuracy when including this term in the linear part far exceeded that of the alternative approach, that is, including it in the nonlinear part. This implies that in order to achieve the best results, one must update the matrix-valued functions containing the linear part at each time step---an equivalent complication to that of the CN method. While this process may be possible in the case of a compact basis representation, such calculations would be infeasible when dealing with large, sparse matrices related to representations such as the real space grid approach. The inclusion of Krylov subspace expansions, or alternative approaches for the evaluation of these matrix-valued functions in such a scenario remains a topic of future research. A split operator approach, using fast Fourier transforms as explained in Sec. \ref{sec:spo}, was tested as a means of approximating the matrix exponential needed for the IFRK4 method. The results of this approach yielded the same improvement of accuracy as those presented above using a diagonalization of the $\Lmat$ matrix. Furthermore, we point out that developing a method based on the formalism described in Appendix \ref{app:lin_time} may provide a means towards improving upon this complication.

\section{Acknowledgment}
This work has been supported by the National Science
Foundation (NSF) under Grants No. Phy 1314463 and
No. IIA126117. This work used the Extreme Science and Engineering Discovery Environment (XSEDE), which is supported by National Science Foundation grant number ACI-1053575.

\appendix
\section{\label{app:lin_time}ETD method with a time-dependent $\Lmat$}
In the case of a time-dependent linear term, $\Lmat(t)$, the general 
form of eq. \eqref{eq:etd_iden} may be written as
\begin{multline}
i\frac{\partial}{\partial t}\left({\rm e}^{i\mathbf{F}(t)}\Psi(\rvec,t)\right)= \\ {\rm e}^{i\mathbf{F}(t)}\left(-\Lmat\Psi(\rvec,t)+i\frac{\partial}{\partial t}\Psi(\rvec,t) \right),
\end{multline}
where 
\begin{equation}\label{eq:F_def}
\mathbf{F}(t) = \int_0^t\Lmat(t')dt'.
\end{equation}
One may, then, use Eq. \eqref{eq:tddft} in order to rewrite this and 
solve for $\Psi(t+\Delta t)$
\begin{multline}\label{eq:etd_td_lin_exact}
\Psi(\rvec,t+\Delta t)={\rm e}^{-i[\mathbf{F}(t+\Delta t)-\mathbf{F}(t)]}\Psi(\rvec,t)+\\i{\rm e}^{-i\mathbf{F}(t+\Delta t)}\int_t^{t+\Delta t}{\rm e}^{i\mathbf{F}(\tau)}\Nmat(\Psi,\tau)d\tau.
\end{multline}
This form is exact for time-dependent $\Lmat$. 

In the limit of small 
$\Delta t$, one may approximate that $\Lmat(t')$ is approximately
constant in the range $t<t'<t+\Delta t$ and that this constant
value is equal to $\Lmat(t)$. This leads to the expression of 
$\mathbf{F}(\tau)$ appearing in the rightmost integral being
\begin{equation}
\mathbf{F}(\tau) = \int_0^\tau\Lmat(t')dt'=L(t)\tau-L(t)t+\mathbf{F}(t).
\end{equation}
Note that the above expression is made possible due to the fact that $\tau$
takes values only between $t$ and $t+\Delta t$ in the context of the integral in eq. \eqref{eq:etd_td_lin_exact}.
One may similarly treat $\mathbf{F}(t+\Delta t)$ appearing in the exponential
factor outside of the integral as
\begin{equation}
\mathbf{F}(t+\Delta t) = \int_0^{t+\Delta t}\Lmat(t')dt'=L(t)(t+\Delta t)-L(t)t+\mathbf{F}(t).
\end{equation}
Lastly, it is simple to show that $\mathbf{F}(t+\Delta t)-\mathbf{F}(t)$ 
in the first term of \eqref{eq:etd_td_lin_exact} is equal to $\Lmat(t) \Delta t$.
After canceling extraneous factors, one is left with eq. \eqref{etd}. 

In calculating the integral in Eq. \eqref{eq:etd_td_lin_exact} one should use time
ordering because $\mathbf{F}(t)$ does not commute for $t$ and $t'$
(the same way as in Eq. \eqref{exact_time_ev_operator}). 
The time-ordering in most calculations in the literature is neglected, assuming that 
if the time step is small enough, the calculation will converge. If
the Hamiltonian is constant in the time interval, then the error of
neglecting the time-ordering is second order in time. For very strong
time-dependent potentials and \rev{fast dependence in the nonlinear part} (dependence on a rapidly changing nonlinear part?),
this may lead to inaccuracies. One possible solution is to use
iterative time-ordering \cite{doi:10.1063/1.3312531}
\begin{multline}
\Psi_k(\mathbf{r},t)={\rm e}^{-i[\mathbf{F}(t)-\mathbf{F}(t_n)]}\Psi_k(\mathbf{r},t_n)\\
-i{\rm e}^{i\mathbf{F}(t)}\int_{t_n}^{t} 
{\rm e}^{i\mathbf{F}(\tau)} \Nmat(\Psi_{k-1},\tau) d \tau.
\label{etdk}
\end{multline}
Using this equation, the time-ordering is forced by iteratively
converging $\Psi_k$ at each time step. This iterative solution
corresponds to the Dyson series and is equivalent
\cite{doi:10.1063/1.3312531} to the Magnus expansion
approach \cite{CPA:CPA3160070404} to  time-ordering. We have checked a
few selected cases and this iteration converges within just one step,
even for large step sizes, so enforcing time-ordering is not necessary in the
calculation.

\section{\label{app:methods}Further Methods}
We are interested in the solution of the initial value problem
\begin{equation}
{d y\over d  t}=f(y,t)\ \ \ \ \ \ y(t=0)=y_0.
\label{de}
\end{equation}

\subsection{\label{app:Euler} Euler method}
Discretizing time with $t_n=n\Delta t$ and defining $y_n=y(t_n)$, the simplest
solution of this equation is given by the (explicit) forward Euler
method
\begin{equation}
y_{n+1}=y_n+\Delta t f(y_{n},t_{n}).
\end{equation}
Alternatively, one can use the implicit backward Euler method
\begin{equation}
y_{n+1}=y_n+\Delta t f(y_{n+1},t_{n+1}).
\end{equation}
Both approaches can be easily derived from first-order Taylor expansions of
$y$, and it is well known that the implicit approach is more stable
but computationally more expensive (one has to determine the
explicitly not known $y_{n+1}$ on the right-hand side).
These approaches are ${\cal O}(\Delta t^2)$ accurate in time.

\subsection{\label{app:RK}Runge--Kutta method}
One may approximate the following step, $y_{n+1}$, from the current step, $y_n$, 
by taking a weighted average of estimated slopes evaluated at temporal 
increments between the two steps. By choosing only one increment, one arrives
at the above described Euler method. For two increments, this is called
the second-order Runge--Kutta method
\begin{align}
y_{n+1}=&y_n+\Delta t k_2, \\ \notag
k_1 =& f(y_{n},t_{n}), \\ \notag
k_2 =& f(y_{n}+\frac{\Delta t}{2}k_1,t_{n+1/2}).
\end{align}

The fourth-order Runge--Kutta method is most widely used, with an error of ${\cal O}(\Delta t^4)$
\begin{align}
y_{n+1}=&y_n+\frac{\Delta t}{6} (k_1+2k_2+2k_3+k_4), \\ \notag
k_1 =& f(y_{n},t_{n}), \\ \notag
k_2 =& f(y_{n}+\frac{\Delta t}{2}k_1,t_{n+1/2}), \\\notag
k_3 =& f(y_{n}+\frac{\Delta t}{2}k_2,t_{n+1/2}), \\\notag
k_4 =& f(y_{n}+\Delta t k_3,t_{n+1}).
\end{align}

\subsection{\label{app:AM} Adams methods}
One can integrate Eq. \eqref{de} as
\begin{equation}
y_{n+1}=y_n+\int_{t_n}^{t_{n+1}} {d y\over d  t} dt
=y_n+\int_{t_n}^{t_{n+1}} f(y,t) dt
\end{equation}
Adams methods approximate the integrand
with a polynomial within the interval $(t_n, t_{n+1})$. Using a 
$k$th-order polynomial, one defines a $(k+1)$th order method. The explicit scheme
is called the Adams--Bashforth method and the implicit one is
called the Adams--Moulton method.

The second-order Adams--Bashforth method can be simply derived by
using a linear interpolation for $f(y,t)$ and is defined as
\begin{equation}
y_{n+1}=y_n+{\Delta t\over 2} \left[3 f(y_n,t_n)-f(y_{n-1},t_{n-1})\right].
\label{ab2}
\end{equation}
The approach is explicit and, as such, it is only conditionally stable---that
is, it requires small time steps. 

The second-order Adams--Moulton method is based on the trapezoidal rule
and is given by 
\begin{equation}
y_{n+1}=y_n+{\Delta t\over 2} \left[f(y_n,t_n)+f(y_{n+1},t_{n+1})\right].
\label{am2}
\end{equation}
This is a stable implicit scheme with longer allowed time steps, but the
trade-off is the  higher
computational cost. These approaches are ${\cal O}(\Delta t^3)$ accurate in time.

\section{\label{app:phi}The $\varphi$-functions}
The $\varphi$-functions are defined as
\begin{equation}
  \label{eq:phifunctions}
    \varphi_0(z) = {\rm e}^z, \ \ \ \ \ 
    \varphi_n(z) = z^{-n}\left({\rm e}^z-\sum_{k=0}^{n-1}{z^k\over k!}\right).
	\end{equation}
The first few functions are
\begin{eqnarray}
\varphi_1(z) &=& \frac{{\rm e}^z - 1}{z}, \quad \\
\varphi_2(z) &=& \frac{{\rm e}^z - 1 - z}{z^2}, \quad \nonumber\\
\varphi_3(z) &=& \frac{{\rm e}^z - 1 - z - \frac{1}{2}z^2}{z^3}.\nonumber
\end{eqnarray}
The $\varphi$-functions satisfy the recurrence relation
	      \begin{equation}
	        \label{eq:phirelation}
		  \varphi_{l}(z) = z \varphi_{l + 1}(z) +
		  \frac{1}{l!}, \qquad
		    \ell = 1, 2, \dots. 
\end{equation}
The efficient and accurate evaluation of these functions is an
important problem that has been addressed in the literature
\cite{doi:10.1137/S0036142995280572,Niesen:2012:A9K:2168773.2168781}.
One major issue is the cancellation error during the direct evaluation
of the $\varphi$-functions \cite{doi:10.1137/S1064827502410633,COX2002430}.	    
Various algorithms have been developed to cope with this problem. The
simplest one is to remove the lowest eigenvalues \cite{COX2002430} if
matrix diagonalization is possible for the calculation of $\varphi$. Another way
is to use a Taylor series
\begin{equation}
  \varphi_n(z) = \sum_{k=n}{z^{k-n} \over k!}.
\end{equation} 
Many more advanced algorithms, including Krylov subspace evaluation
\cite{doi:10.1137/140964655,doi:10.1137/S0036142995280572} and other methods used to calculate matrix
exponentials \cite{doi:10.1137/100788860} have also been developed.

\end{document}